\documentclass{aa}
\usepackage{graphicx}
\usepackage{txfonts}

\usepackage{diagbox}
\usepackage{rotating}
\usepackage{tikz}
\usepackage{multirow}
\usepackage{wasysym}
\usepackage{adjustbox}
\usepackage{afterpage}
\usepackage{placeins}
\usepackage{xcolor}
\usepackage{caption}
\usepackage[strict]{changepage}
\usepackage{lscape}
\usepackage{pdflscape}
\usepackage{soul}

\usepackage{hyperref}
\hypersetup{colorlinks=True,citecolor=blue}

\def\ms{\hbox{\,m\,s$^{-1}$} }         
\def\cms{\hbox{\,cm\,s$^{-1}$} }       
\def\m2s2{\hbox{\,m$^{2}$\,s$^{-2}$} } 
\def\kms{\hbox{\,km\,s$^{-1}$} }       
\def\vsini{\hbox{$v$\,sin\,$i$}}      
\def\Mearth{\hbox{$\mathrm{M}_{\oplus}$}}             
\def\Msun{\hbox{$\mathrm{M}_{\odot}$}}             

\def\snr{S/N$_{cont}$}

\def\logrhk{$\log$(R$^{\prime}_{HK}$) }
\def\ang{\text{\AA}}

\usepackage{amstext}

\begin{document}

   \title{Stellar activity correction using PCA decomposition of shells}
   


   \author{M. Cretignier\inst{1}
          \and X. Dumusque \inst{1}
          \and F. Pepe \inst{1}
          }

   \institute{Astronomy Department of the University of Geneva, 51 ch. de Pegasi, 1290 Versoix, Switzerland\\
              \email{michael.cretignier@unige.ch}
             }

   \date{Received XXX ; accepted XXX}

 
  \abstract
   {Stellar activity and instrumental signals are the main limitations to the detection of Earth-like planets using the radial-velocity (RV) technique. Recent studies show that the key to mitigating those perturbing effects might reside in analysing the spectra themselves, rather than the RV time series and a few activity proxies.}
   {The goal of this paper is to demonstrate that we can reach further improvement in RV precision by performing a principal component analysis (PCA) decomposition of the shell time series, with the shell as the projection of a spectrum onto the space-normalised flux versus flux gradient.}
   {By performing a PCA decomposition of shell time series, it is possible to obtain a basis of first-order spectral variations that are not related to Keplerian motion. The time coefficients associated with this basis can then be used to correct for non-Dopplerian signatures in RVs.}
   {We applied this new method on the YARARA post-processed spectra time series of \object{HD10700} (\object{$\tau\,Ceti$}) and \object{HD128621} (\object{$\alpha\,Cen\,B$}). On HD10700, we demonstrate, thanks to planetary signal injections, that this new approach can successfully disentangle real Dopplerian signals from instrumental systematics. The application of this new methodology on HD128621 shows that the strong stellar activity signal seen at the stellar rotational period and one-year aliases becomes insignificant in a periodogram analysis. The RV root mean square on the five-year data is reduced from 2.44 \ms down to 1.73 \ms{}. This new approach allows us to strongly mitigate stellar activity, however, noise injections tests indicate that rather high signal-to-noise ratio (S/N$ > 250$) is required to correct for the observed activity signal on HD128621.}
   {}

   \keywords{methods:data analysis -- techniques:radial velocities -- techniques:spectroscopic -- star individuals: HD128621}

   \maketitle

\section{Introduction}

The detection of Earth-like exoplanets orbiting Sun-like stars remains one of the most exciting perspectives for the future of astrophysics and one of the most tremendous challenges for the next few years. Such detections have so far been out of reach of the radial velocity (RV) technique, as the most precise spectrographs, HARPS \citep{Mayor(2003)}, HARPS-N \citep{Cosentino(2012)}, and HIRES, have been reaching a precision of $\sim$1 \ms{}. This is an order of magnitude greater than the 0.1 \ms{} RV semi-amplitude with respect the Earth and the Sun.

Despite the technical challenges to reach extreme precision, the RV method remains the most promising technique for the detection of other Earths due to the low transit probability of those objects and the extremely dim light emitted or reflected by their surface \citep{Zhu(2021)}. This has motivated the design of new ultra-stable spectrographs such as ESPRESSO \citep[][]{Pepe(2021)} and EXPRES \citep[][]{Jurgenson:2016aa} that already demonstrated a 30 \citep[][]{Suarez-Mascareno:2020aa} and 60 \cms \citep[][]{Brewer:2020aa} RV precision, respectively.

Even if the technical challenge in term of RV precision has been largely overcome\footnote{this still has to be proven on the long-term}, at this level of precision, stellar signals dominate the RV budget, thus making the detection of Earth-like planets around Sun-like stars extremely difficult. The problem is complex since stellar signals contaminate the RVs in multiple ways and RV measurements often do not contain any simultaneous observables, such as photometry or imaging of the stellar surface (as for the Sun), which could help in constraining the problem \citep[e.g.][]{Milbourne(2021), Milbourne(2019)}. Thankfully, stellar signals  affect high-resolution spectra in a different way than a planetary signal. Indeed, while a Keplerian signal affects all the lines in the same way, stellar activity affects differently each individual spectral line \citep[e.g.][]{Thompson(2017),Wise(2018),Dumusque(2018)}, which offers a pathway to solving  this problem \citep[e.g.][]{Davis(2017)}.

Stellar signals are, nonetheless, challenging to mitigate, as they are: i) acting on several timescales; ii)  semi-periodic; iii)  stellar-type dependent; iv)  driven by different physical processes; and v)  modifying  each spectral line in a different way. Stellar signals introduce short-terms RV variations through stellar oscillations \citep[e.g.][]{Dumusque(2011)a} as well as minutes from day perturbations due to granulation and supergranulation, rotational semi-periodic modulations due to the evolution of active regions on the stellar surface, and long-term trends due to stellar magnetic cycles \citep[e.g.][]{Dumusque(2011)b, Meunier(2010)}. 

The most important physical processes contributing to the RV budget are the convective blueshift inhibition (CBI) \citep{Meunier(2010), Cretignier(2020a)}, the flux contrast breaking the flux balance between approaching and receding stellar limbs \citep[e.g.][]{Saar-1997b,Dumusque(2014),Donati(2017)}, 
and line-strength variations due to temperature fluctuations \citep{Wise(2018),Thompson(2017),Basri(1989)}. Even if the later perturbation of stellar line is assumed to be symmetric and would not therefore induce a Doppler shift, it does so for blended lines \citep[e.g.][]{Cretignier(2020a)}. All those important perturbing effects are induced by stellar magnetic activity and are often referred to as stellar activity. 

In this paper, we first discuss in Sect.~\ref{sec:CCF}, the main limitations of the use of the cross-correlation function \citep[CCF,][]{Baranne(1996)} as a high signal-to-noise ratio (S/N) representation of a spectrum to correct for stellar activity signal. We then describe in Sect.~\ref{sec:folding} a new spectral representation aimed at 
overcoming those limitations and we discuss in Sect.~\ref{sec:shells} how we can use this representation to mitigate the stellar and instrumental signals. We then apply this new approach to HD10700 ($\tau\,Ceti$) and HD128621 ($\alpha\,Cen\,B$) and we show the results in Sect.~\ref{sec:planets_injections} and Sect.~\ref{sec:activity}, respectively. We finally investigate the minimum signal-to-noise ratio (S/N) of the spectrum required by this new approach to be efficient in Sect.~\ref{sec:noise_sim} before presenting our conclusions in Sect.~\ref{sec:conclusion}.

\section{Disadvantages of analysing spectra or cross-correlation function time series
\label{sec:CCF}}

\citet{Davis(2017)} showed that PCA could theoretically be used to disentangle stellar activity from planetary signals at the spectra level. However, the main limitations of PCA is generally brought on by the presence of outliers and the low S/N of the training dataset. Thanks to YARARA post-processing (see Sect.\ref{sec:processing}), the majority of outliers should be corrected for at the spectrum level, leaving the problem of the low S/N to be solved. To increase the S/N, a commonly used technique is to cross-correlate the stellar spectrum with a mask that contains holes at the position of each main spectral line. This is mathematically equivalent to performing a weighted average of the stellar lines in the velocity space.
The obtained product, known as the CCF, is an averaged line profile with the highest possible S/N.

Two promising methods to model non-Dopplerian signals at the CCF level have been developed recently. Firstly, the SCALPEL algorithm \citep[][]{Collier(2021)} which projects the CCF in the auto-correlation function space known to be shift-invariant. Hence, a Doppler-invariant average line profile is obtained, followed by a modelling of the variations seen in this line profile using a PCA. Results on the HARPS-N solar data set \citep[][]{Dumusque(2021)} show that planetary signals with amplitudes as small as 0.4 \ms{} can be recovered. By using a convolution neural network (CNN) to model the variations seen at the CCF level, \citet{Beurs:2020aa} show that an 3.5 Earth-mass planet at 365 days, with an amplitude of 0.3 \ms{}, could be recovered on the same solar data set. However, \citet{Beurs:2020aa} inject the planetary signal after correcting for the CCF variations using the CNN, and it is not clear if the CNN would absorb such a signal. Even if this is the case, a solution could be found by fitting simultaneously Keplerian signals along with the CNN. Although the techniques described above using the CCF seem promising to mitigate non-Dopplerian signals, the CCF representation presents three main disadvantages, as follows. 

First, the CCF is "white", meaning that line profiles from different wavelengths are averaged-out together, which prevent the detection of the colour dependence of stellar activity \citep[e.g.][]{Huelamo(2008)}. A possible solution to solving this problem consists in computing order-by-order CCFs \citep[e.g.][]{Zechmeister(2020)}, however, the wavelength limits are somewhat arbitrary and often based on instrumental considerations. 

Secondly, the CCF is a global average line profile. As spectral lines are weighted by their depth \citep[][]{Pepe(2002)}, deep and shallow lines are both projected onto the same line profile. As a consequence, it is not possible to detect the line depth effects that are expected to play a major role in characterising the CBI \citep{Cretignier(2020a)}. This problem could be overcome by deriving several CCFs using different masks that group lines by their depth. However, the production of such masks remains exhausting and non trivial in particular when combined with the wavelength effect mentioned here above.
    
Lastly, the definition of the RV is unclear in the CCF space. Strictly speaking, if the absolute position of the stellar lines were known, the RV value should be defined as the minimum (or maximum depending on the definition of the cross correlation mask) of the CCF. Unfortunately, laboratory wavelengths often differ from stellar lines positions due to blends, but also due to the convective blueshift \citep{Reiners(2016),Lohner(2019)} that skew stellar line profiles \citep{Saar(2009)} and thus prevent the detection of an accurate stellar RV. It is, however, possible to measure a precise RVs by fitting the CCF profile using an empirical model that can be a parabola fit on the CCF core, a Gaussian fit, a Voigt profile fit, a Bi-Gaussian fit \citep[][]{Figueira(2013)}, or a skew Normal \citep[][]{Simola:2019aa}. None of those models are perfect as each of them present some advantages and drawbacks, making the choice of the model rather arbitrary in the end.

Any method that works with CCFs can not simultaneously deal with all the issues listed above in a straightforward and easy way thanks to the averaging nature of the cross-correlation operation. We propose below a new possible representation of a stellar spectrum, halfway between the spectrum and the CCF in terms of complexity, solving for the different disadvantages of the CCF discussed above.

\section{Method}
\label{sec:method}

\subsection{Data pre-processing
\label{sec:processing}}

We worked with HARPS 1D-merged spectra produced by the official data reduction software (DRS). The spectra were post-processed using the YARARA pipeline \citep{Cretignier(2021)} to remove the known systematics present on HARPS spectra (instrumental and telluric contamination). In YARARA, the 1D-merged spectra are first nightly stacked, before being continuum normalised by RASSINE \citep{Cretignier(2020b)}. The nightly-binned normalised spectra time series is then corrected for known systemics using a succession of algorithms dedicated to correct all the systematics. In order of processing, YARARA performs correction for i) cosmics; ii) tellurics; iii) interference pattern; iv) ghosts; v) stitchings; and vi) fibre B contamination. To derive more accurate line-by-line (LBL) RVs \citep{Dumusque(2018)}, a data-driven line selection was performed for each star, as in \citet{Cretignier(2020a)}. The residual contaminations after YARARA processing are mainly stellar activity variations and instrumental systematics undetectable on a single line profile due to low S/N. The following sections explain how RVs can be further corrected for systematics thanks to an analysis of spectra time series in a new dimensional space.

\subsection{Projection of the spectra onto a new dimensional space
\label{sec:folding}}

In the case of a small velocity shift, $\delta v,$ with respect to the spectrum sampling, we can show via linearisation (first order of a Taylor series) that a shifted spectrum, $f,$ can be expressed as the sum of a non-shifted spectrum, $f_0,$ and of the gradient of the flux multiplied by the difference in wavelength $\lambda - \lambda_0$ for the same point in the shifted and non-shifted spectra \citep{Bouchy(2001)}:
\begin{equation}
\label{eq:0}
f(\lambda) = f_0(\lambda_0)+\frac{d f_0}{d \lambda} (\lambda - \lambda_0)+\mathcal{O}((\lambda-\lambda_0)^2).
\end{equation}
Using this formula and the fact that a Doppler shift can simply be expressed by $\delta v/c = \delta \lambda/\lambda$, we can show that:
\begin{equation}
\label{eq:1}
\delta v \simeq \frac{f-f_0}{\lambda}\cdot c \cdot \left(\frac{d f_0}{d \lambda}\right)^{-1} \equiv \delta f(\lambda) \cdot \left(\frac{d f_0}{d \lambda}\right)^{-1}
,\end{equation}
where $\delta v$ is the Doppler shift and $c$ the speed of light in vacuum. Thus, a velocity shift in that approximation regime is simply the linear coefficient between the flux variation (divided by the wavelength) and the flux gradient:

\begin{equation}
\label{eq:2}
\delta f \left(f_0,\frac{d f_0}{d \lambda}\right) = \delta v \cdot \left(\frac{d f_0}{d \lambda}\right) \equiv \delta v \cdot DS \left(f_0,\frac{d f_0}{d \lambda}\right).
\end{equation}
Here $DS$ stands for Doppler shell basis component and it represents, in the shell space, the behaviour of a pure Doppler shift. This denomination will become more clear in Sect.~\ref{sec:shells}.

\begin{figure}[t]
        \centering
        \includegraphics[width=9cm]{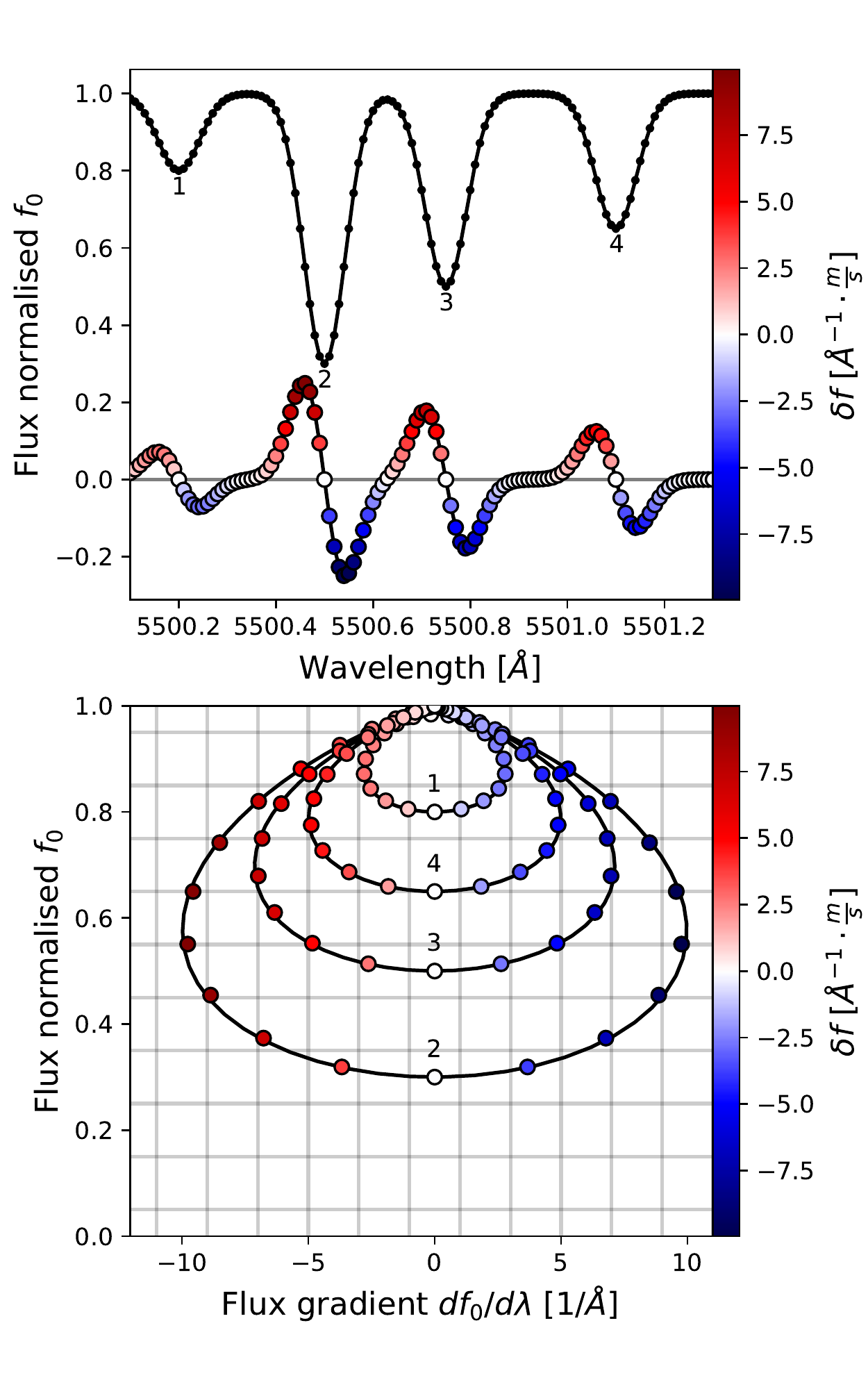}
        \caption{Projection of a small part of a spectrum into the $(d f_0/d \lambda,f,\,\delta f)$ space that we call the shell space. \textbf{Top:}  Reference spectrum used to perform the change of variables (in black). The coloured dots correspond to the value of $\delta f$ when comparing a Doppler shifted spectrum to the reference spectrum. \textbf{Bottom:} Projection of the Doppler shifted spectrum onto the shell space. Each point in this new space corresponds to an original point of the reference spectrum and the colour coding corresponds, as above, to the value of $\delta f$ when comparing a Doppler shifted spectrum to the reference spectrum. Positions of the lines' cores were indexed to make the identification easier. The mean velocity difference between the shifted and reference spectra $\delta v$ can be recovered as the slope between the flux gradient and $\delta f$.}
        \label{FigShell}
\end{figure}

\begin{figure}[t]
        \centering
        \includegraphics[width=9cm]{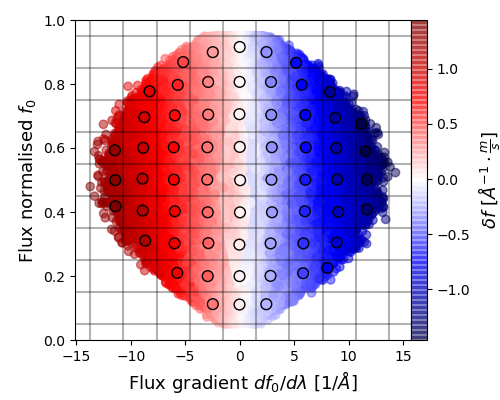}
        \caption{Projection of the Doppler shifted spectrum onto the shell space for a real spectrum of HD10700. The shell space has been divided in a $9\times 9$ grid (black lines) to increase the S/N of the raw shell after binning all the points within each cell.}
        \label{FigShell3}
\end{figure}

This finding leads to the observation that if we want to extract a velocity value, the gradient of the flux appears as a much wiser variable than the wavelength to characterise a spectrum, $S$. Moreover, keeping an information on the flux is primordial if we want to probe stellar activity signatures that are known to modify the line strength. Finally, to compute the velocity difference between a spectrum and a reference spectrum, we need to consider the observable $\delta f$. Therefore, as shown in Fig.~\ref{FigShell}, we decided to represent a spectrum in the space $(d f_0/d \lambda,\,f)$ rather than the common one $(\lambda,\,f)$. In addition, we added the extra dimension, $\delta f$ (shown as a colour scale in Fig.~\ref{FigShell}) to be able to compute the RV difference between a given spectrum and a reference one. This transformation produces a compact representation of the spectrum where identical line profiles follow the same loop path. Hereafter, we refer to the $(d f_0/d \lambda,\,f,\,\delta f)$ space as the shell space and the projection of a spectrum onto this space $S(d f_0/d \lambda,\,f,\,\delta f)$ as a raw shell, due to the resulting shape of a spectrum in this newly defined coordinate system. In the shell space, the core of the lines are located at a gradient of zero and a flux level smaller than 1, whereas all spectral points from the continuum are projected onto the same single point with flux unity and a null gradient. An example of the full spectrum of  the G8 dwarf HD10700 is displayed in Fig.~\ref{FigShell3} for a real Doppler shift.

\begin{figure}[t]
        
        \centering
        \includegraphics[width=9cm]{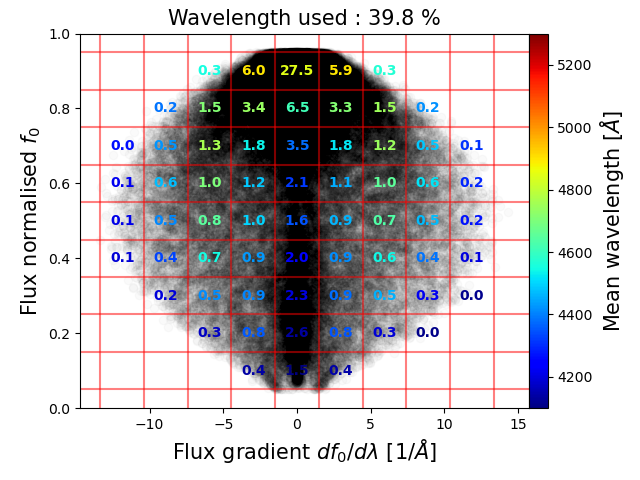}
        \caption{Shell representation and statistics of a spectrum from HD10700. Here, we highlight the fraction of spectral elements that fall within each bin, as well as their average wavelength (in colour scale).}
        \label{FigShell2}
        
\end{figure}

To increase the S/N of a raw shell, we can bin the data in the shell space following a grid in normalised flux and flux gradient. Hereafter, we will call this binned raw shell simply shell. We note that the position of a bin in the $(d f_0/d \lambda,\,f)$ space is the barycenter of all points in each cell, and its $\delta f$ value is obtained by performing a weighted average on the $\delta f$ of all the points within a bin, considering as the weight the inverse squared of the uncertainty in $\delta f$. For each bin, we calculate the corresponding S/N\footnote{In this case, the S/N is defined as the ratio between 1.48 times the median absolute deviation (MAD) of all points within a bin (equivalent to a standard deviation, but less sensitive to outliers) and the median uncertainty in $\delta f$.} and only keep bins for which the S/N is larger than 1, to prevent being affected by noise occurring at the limb of a shell. We also discard the normalised flux level above 0.95 and below 0.05, since those regions do not contain significant Doppler information. In Fig.~\ref{FigShell2}, we can see for HD10700 the fraction of spectral points, as well as their mean wavelength, which fall into each bin of the raw shell defined by a $9\times9$ grid. We highlight that only 39.8\% of the spectrum is used, since the remaining part is made of the stellar continuum which is discarded.

Using this shell representation, we can observe that the relevant parts of the spectrum that are sensitive to Doppler shift, which are the part with the highest absolute flux gradient value, represent a very small fraction of the total spectrum. Indeed, the bins with a value $|d f_0/d \lambda|>4.4$ only represent $39.8\%\cdot18.1\%=7.2\%$ of the total spectrum. Doing a similar analysis for the K1 dwarf HD128621 leads to a similar value of : $60.4\%\cdot15.3\%=9.2\%$. This observation explains why localised systematics can strongly affect the overall stellar RV even if the contamination appears in a small portion of the spectrum \citep{Cretignier(2021)}.

Finally, we observe that the positioning of spectral points in the shell space follow a colour dependence, which demonstrate that, opposite to a 'white' CCF, the shell representation can be sensitive to chromatic effects. However, the small fraction of stellar lines in the red spectral range produces a reduced wavelength span compare to the full span of the spectrum, with average values ranging between 4200 and 5200 $\AA$ (see colourbar of Fig.~\ref{FigShell2}). For this reason, shells are more dedicated to disentangle line depth and asymmetric line shape variations than chromatic ones.

The shell representation presented in this section allows us to solve most of the issues mentioned in Sect.~\ref{sec:CCF} when analysing spectra or CCF time series. 
For an analysis of spectral variations using PCA, this new representation is more appropriate than the spectra themselves as it provides data with a much higher S/N. This new representation also contain more information than a CCF as: i) red lines, that are broader than blues lines\footnote{Also, the core of blue lines tend to reach lower flux values than red lines because of the larger line density in the blue spectral range which is inducing line blending} for a fixed velocity width and similar depth, will follow different loop paths; ii) deep lines and shallow lines can be differentiated; and iii) the velocity $\delta v$ can simply be obtained by performing a linear least-square to estimate the linear coefficient between $\delta f$ and the flux gradient. We note however two remaining issues. First, some degeneracies brought by line blending, which could be partially lift off if the second derivative of the flux was also included, however, this would lead to region of the ($f$,$f'$,$f''$) space with low S/N which is not desired (see already the low fraction of spectral point per bin in Fig.~\ref{FigShell2}). Secondly, we point out the necessity of obtaining a reference spectrum, which is not straightforward. Such a spectrum can be obtained by stacking spectra in the stellar rest-frame after subtracting a first guess for the RVs. This method was in fact already implemented in YARARA and represents the main philosophy of the pipeline, where a master spectrum is constructed directly from the observations. 

\subsection{Decomposition of shells onto an orthogonal basis.
\label{sec:shells}}

In the previous section, we demonstrated the behaviour of a pure Doppler-shift on a shell. We recall that a shell is a spectrum projected onto the shell space, and then binned onto a $9\times 9$ grid defined in the gradient flux and normalised flux dimensions. However, stellar activity, by inducing line-width broadening, line-depth, and bisector variations, will modify the shell in a different way than a pure Doppler shift would. Thus, if we could find an orthogonal 2D basis that describes a shell, we should be able to separate a pure Doppler-shift variation from a more complex perturbation induced by stellar activity or instrumental signals.

In order to separate Dopplerian components from other systematics, we have to define a basis of 2D functions that will be fitted on the shells, such basis will hereafter be called  the 'shells basis'. Those 2D functions have to be orthogonal with each other and should fit the variance of shells as much as possible. A PCA is therefore well suited to provide the optimal shell basis. However, the shell basis components derived by the PCA, $PS$, will not be orthogonal to the Doppler shell basis component. We thus used a Gram-Schmidt algorithm to guaranty orthogonality between the PS and $DS$ shell basis components. With such a basis,  a shell at time $t$ can be written as : 
\begin{equation}
\label{eq:3}
\delta f \left(f_0,\frac{d f_0}{d \lambda},t\right) = \alpha_{DS}(t) \cdot DS\left(f_0,\frac{d f_0}{d \lambda}\right) + \sum^N_{j=1} \alpha_{j}(t)\cdot PS_j\left(f_0,\frac{d f_0}{d \lambda}\right) 
.\end{equation}

Looking at the equation above, it may seem that several options are on hand to correct for non-Dopplerians signatures, but the fact is that only one is well suited for the task. For instance, Eq.~\ref{eq:3} cannot be used to correct the spectra in flux since no bijection exist from shells to spectra 
due to the mentioned degeneracy induced by blended lines. Another solution would consist of simply extracting the $\alpha_{DS}(t)$ coefficient related to the Doppler shell (equivalent to $\delta v$ in Eq.~\ref{eq:2}), somewhat equivalent to a template matching method. However, this coefficient is generally almost identical to the CCF RV time series and therefore does not improve the RVs strongly. This can be understood, since no fundamental reason can justify why both stellar activity and instrumental systematics could not present a real Doppler-shifted component affecting as well $\alpha_{DS}(t)$ (see Sect.~\ref{sec:activity1}). However, it is really unlikely for those contaminations to solely provide a Dopplerian component without presenting simultaneously non-Dopplerian effects \citep{Huelamo(2008),Meunier(2010),Reiners(2013),Marchwinski(2015),Fischer(2016),Davis(2017),Barnes(2017),Zechmeister(2018),Dumusque(2018),Cretignier(2020a),Lisogorskyi(2021),Lienhard(2021)} that would be modelled by the PCA shell basis components. As a consequence, the best option remains to decorrelated in the time-domain $\alpha_{DS}(t)$ with the other coefficients $\alpha(t)$, an option also chosen by the SCALPEL method \citep{Collier(2021)}.

Instead of correcting $\alpha_{DS}(t)$ or the CCF RVs, we chose to correct the LBL RVs by adding the $\alpha_j(t)$ shell coefficients in the multi-linear model that will be fitted:
\begin{equation}
\label{eq:5}
\widehat{RV}_i(t) = RV_i(t) - \sum^N_{j=1} \gamma_{i,j}\cdot \alpha_j(t) \equiv \widehat{RV}_i (\alpha_N).
\end{equation}
A similar result would be obtain by decorrelating the CCF RVs, but the study of the $\gamma_{i,j}$ can provide deeper information on the nature of the fitted components, in order to assess the origin of a component (from stellar activity or instrumental systematics). By averaging as in \citet{Dumusque(2018)}, the LBL RVs obtained in \citet{Cretignier(2021)} produce the average $RV_m(t)$ time series that will be used in this work.

 \begin{figure*}[tp]
        \centering
        \includegraphics[width=18cm]{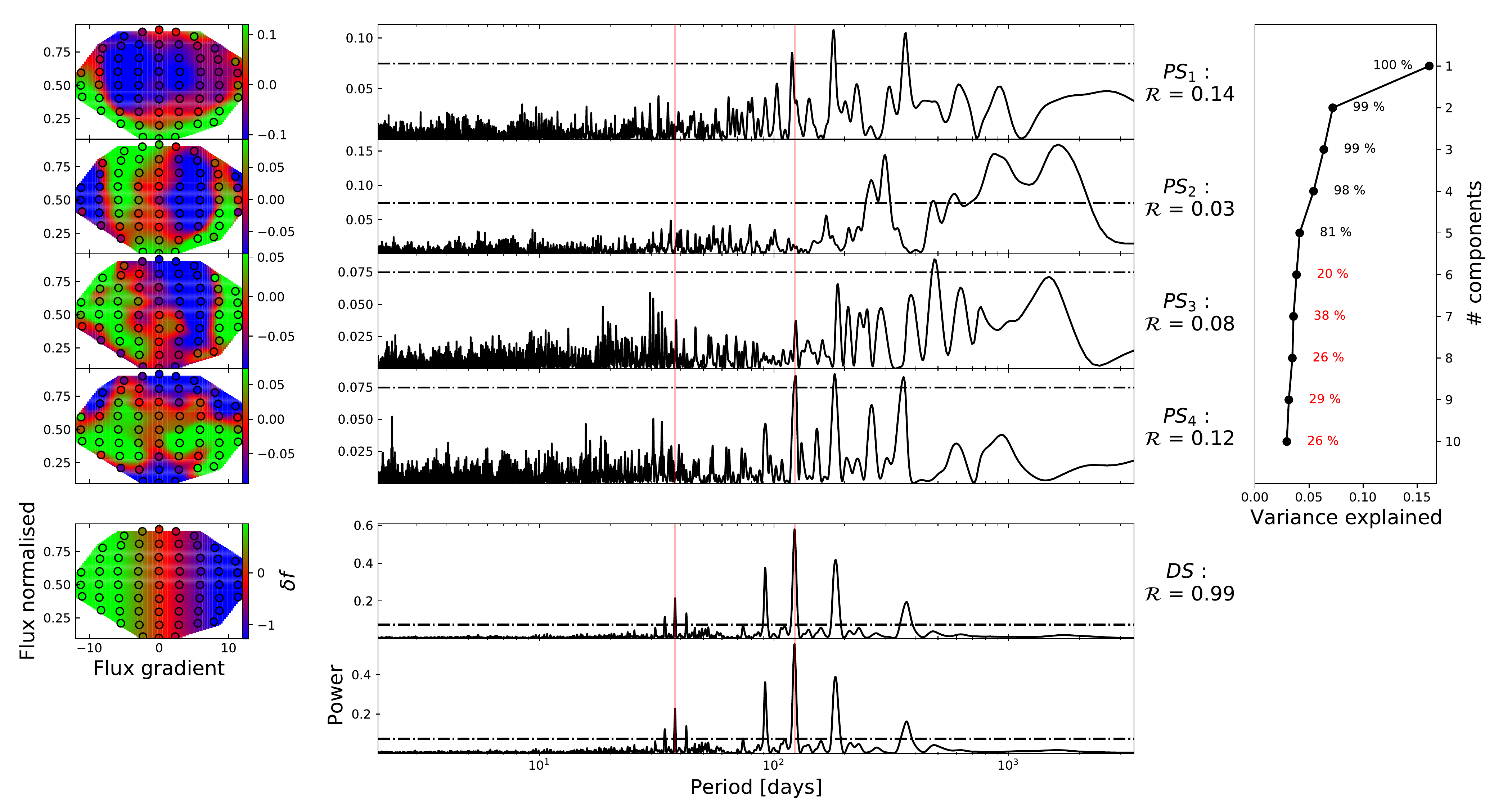}
        \caption{ Shell decomposition of HD10700 spectra. \textbf{Left: } Representation of the different shells basis components and the time series of the coefficients $\alpha_j(t)$ in front each of those shell basis components when modelling the RVs of HD10700 (see Eq.~\ref{eq:3}). All the components are orthogonal to the Doppler shell basis component ($DS$) displayed at the bottom (similar to Fig.~\ref{FigShell3}) thanks to a Gram-Schimdt algorithm. The pixel grid in the shell space on which the PCA was trained (dots) is shown, as well as a smooth function interpolated on the full shell space using a cubic interpolation algorithm. \textbf{Middle: } Generalised Lomb-Scargle periodogram of the $\alpha_j(t)$ shell component coefficients. The 0.1\% false alarm probability (FAP) level for each periodogram is indicated by an horizontal dotted-dashed line. From top to bottom, we can see the five first principal shell basis components and the $DS$. The Pearson correlation coefficient between the $\alpha_j(t)$ and the CCF RV time series is indicated on the right side of each subplot. The lowest periodogram represents the RV time series linearly decorrelated from the $\alpha_j$ coefficients (similar to Eq.~\ref{eq:5}). 
        The power at the planetary injected periods (vertical red lines) is never the dominant power for any shell. This is particularly true for the 37-day signal less significant than the 0.1\% FAP level. Some power is observed around 122 days for $PS_1$ and $PS_4$ due to an alias of a one-year systematics (see main text). \textbf{Right: } Explained variance curve of the PCA components.
        Our cross-validation algorithm (see Appendix.~\ref{appendix:a}) provides four significant components ($>95\%$). The components compatible with a single outlier explanation ($<80\%$) are shown in red.}
        \label{FigShells2}
\end{figure*} 

Some concern may be raised about the number of PCA shell basis components that  have to be fit. This is not a simple question to answer, as the number of relevant components change as a function of the S/N and with the total number of observations. 
Fitting too many components that are mainly noise-dominated would solely result in an undesirable noise increase in the fitted model or an overfitting. On the contrary, under-fitting would imperfectly correct for the contaminations. Therefore, a trade-off has to be found which can be accomplished by performing cross-validation (see Annex.~\ref{appendix:a}). In summary, our cross-validation algorithm measures the occurrence rate of PCA shell basis components by selecting random subsets of observations. For the present paper, we performed 300 independent sub-selections of the shells time series by randomly removing  20\% of the observations. In the end, we only considered the PCA shell basis components having an occurrence rate higher than 95\% as significant.

\section{Results
\label{sec:results}}

In this section, we demonstrate the applications of the shell decomposition presented in Sec.~\ref{sec:shells} 
on two HARPS datasets that present high S/N and an exceptional sampling. We first analyse the data of HD10700 that present no sign of stellar activity and no
significant planetary signals (see however the very small amplitude candidates published by \citet{Feng(2017)}). This dataset will be used to prove, similarly to what was done for YARARA in \citet{Cretignier(2021)}, that our shell approach does not absorb any injected planetary signals with amplitudes smaller than 3 m/s. We then analyse the data of HD128621 to show that our shell approach can significantly correct for stellar activity. Test of planetary signal injection into the HD128621 dataset also show that in that case, the planetary signal is fully recovered despite strong stellar activity modulations in RV. Finally, we estimate the S/N that is required to mitigate the stellar activity signal in HD128621 using our shell approach.

\subsection{Planetary injections on HD10700
\label{sec:planets_injections}}

HD10700 ($\tau\,Ceti$) is a bright G8V star of magnitude $m_v=3.5$ with a peculiarly low activity level. Opposite to 
HD128621 who shows clear evidence of stellar activity manifestation, HD10700 has never exhibit any strong evidence of activity over the two decade of HARPS observation and is likely the quietest star in the solar vicinity. Such unusual behaviour could be explained either by a perturbation of the stellar magnetic cycle similar to the solar Maunder minimum or by the star being seen pole-on. As a consequence, HD10700 has been intensively observed as a standard calibration star during the full lifetime of HARPS in order to track instrumental systematics and search for very low-mass planets.

Here, we analyse  ten years of the HD10700 HARPS dataset gathered before the fibre upgrade in June 2015. In \citet{Cretignier(2021)}, we demonstrated that by post-processing the spectra with YARARA, which removes most of the known instrumental systematics, a RV precision of 1 \ms{} can be reached over those ten years. To perform a planetary signal recovery test, we used the same dataset as in that paper, since the injected planets -- with 3 and 2 \ms{} signals at 122.4 and 37.9 days, respectively -- were dominating the RV variation. As in \citet{Cretignier(2021)}, the planets were injected at the spectrum level by first correcting the systematics, Doppler-shifting the cleaned spectra with the Keplerian solutions, and finally re-introducing the removed systematics before re-running YARARA. 

On the left of Fig.~\ref{FigShells2}, we plot the shell basis components (the first PCA basis components in addition to the $DS$ basis component) that we obtained after applying the shell decomposition presented in Sect.~\ref{sec:shells}. Once the optimal basis is found, each shell that is obtained after projecting each stellar spectrum onto the shell space and binning it is expressed as a linear combination of the different $j$ components of the basis (see Eq.~\ref{eq:3}) and the related coefficient $\alpha_j(t)$ are saved. For each basis component, $j$, we therefore obtained a time series $\alpha_j(t)$ and we show the plot of its corresponding Generalised Lomb-Scargle (GLS) periodogram in the middle of Fig.~\ref{FigShells2}. All the false alarm probability (FAP) used in this paper for the different periodograms were obtained by bootstrap of 10'000 random permutations. On the right of Fig.~\ref{FigShells2}, we show the results of our cross-validation analysis, described in Appendix.~\ref{appendix:a}, which estimates whether the shell basis components are significant or not. This analysis shows that the first four shell basis components are found significant (the matrix of the cross-validation algorithm can be found in Fig.~\ref{FigCV2}). 
Finally, in the last subplot of Fig.~\ref{FigShells2}, the periodogram at the very bottom of the figure corresponds to the GLS periodogram of the residuals, $\widehat{RV}_m$ (see Eq.~\ref{eq:5}), after all the significant PCA shell basis components have been fitted.

By looking at the shell basis component in Fig.~\ref{FigShells2}, we clearly see that all  shells present a clear and smooth structure. For the PCA algorithm, each point of the shell is considered independent, and thus the algorithm has no constraints that would produce a smooth shell except if the physical process perturbing the line profile is itself a smooth perturbation. This is, of course, something that is expected from both stellar activity and instrumental systematics.



We clearly see in Fig.~\ref{FigShells2} that the 37.9 and 122.4-day planetary signals are completely captured by the $DS$ shell basis component. Other peaks surrounding the periods of the injected signals are the one-year aliases (see e.g \citet{Fischer(2016),VanderPlas(2018)}) due to the window function produced by the gap between the observational seasons. We also observe some power around 122 days for the first and fourth components. This period is not the dominant signal of respective periodograms and does not correspond to the 122.4-day injected planet. This power seems to be produced by the window function of a one-year systematic corrected. Indeed, it can be shown that by generating a perfect one-year signal with the time sampling of the observations, the GLS periodogram clearly presents significant power at one-year, but it also does so at its first two  harmonics as well.

The time series of the coefficient $\alpha_{DS}(t)$ fitted for this shell component correlates perfectly with the classical CCF RVs ($\mathcal{R}=0.99$). Despite those planetary signals being strong in the $DS$ shell basis component, we investigated by how much the planetary signals were affected by the linear decorrelation with the $\alpha_j$ shell basis coefficients. We thus fitted a non-eccentric two-Keplerian solution on the residuals $\widehat{RV_m}$ by fixing the periods of the injected planets, and we recovered amplitudes of 2.83 \ms{} and 2.00 \ms{} for the 3 and 2 \ms{} injected planetary signals, respectively. We note that the period of the planet after YARARA post-processing was in fact already recovered at 2.85 m/s \citep{Cretignier(2021)}. This was perfectly explicable by the known presence of lower amplitudes signals around this star \citep{Feng(2017)}. By fitting all signals more significant than a FAP of 0.1\% , the planetary amplitudes change to 2.95 and 2.03 \ms{}, respectively, indicating some cross-talk between the injected planets and pre-existing lower amplitude signals. 

\begin{figure*}[tp]
        \centering
        \includegraphics[width=18cm]{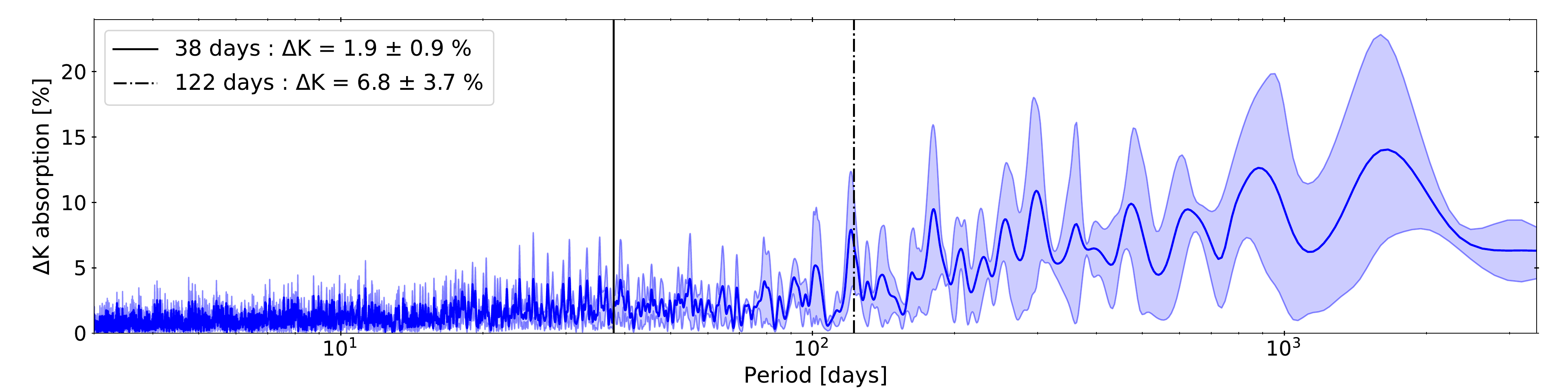}
        \caption{Measurement of the collinearity between planets in circular orbits and the $\alpha_j(t)$ basis for HD10700. The simulations are made of 18 different phases and 15'000 different periods. The envelope describes the 1-sigma dispersion obtained from the different phases for the same period. The injected planets are marked by vertical lines. A typical decrease of $2\pm1$ and $7\pm4$\% can be expected for the 37-day and 122-day planets, respectively. Those results are coherent with the $\sim 2 \ms{}$ and $\sim 2.85$ \ms{} planetary signals recovered, while the original signals injected for those planets was 2 and 3 \ms{}, respectively.}
        \label{FigAbs}
\end{figure*} 

As mentioned in Sect.~\ref{sec:shells}, only the PCA shell basis components are orthogonal to the $DS$ shell basis component, but not the $\alpha_j$ coefficients with respect to $\alpha_{DS}(t)$. 
As a consequence, all planetary signals would be slightly projected on the basis of the $\alpha_j(t)$ coefficients, and the best way to account for this is to include the planetary signals in the multi-linear regression and perform a Monte Carlo Markov Chain to measure the degeneracy. We note however that such projection is reduced when the number of observations increases and nearly vanishes for HD10700 due to the large number of observations. 

Assessing the question of the minimal number of observations required is difficult to answer. First, because the observations should probe different 'levels of contamination' in order to allow the PCA to detect the flux variations and, secondly, since with lower number of observations, the degeneracies of the full Keplerian model increases, which itself depends on: the model complexity, the period of the planetary signals, and power spectrum of the $\alpha_j(t)$ basis. Despite the difficulty in estimating a number on the basis of too many unknown parameters, there is a simple way to estimate the degeneracy.

\begin{figure*}[h!]
        
        \centering
        \includegraphics[width=18cm]{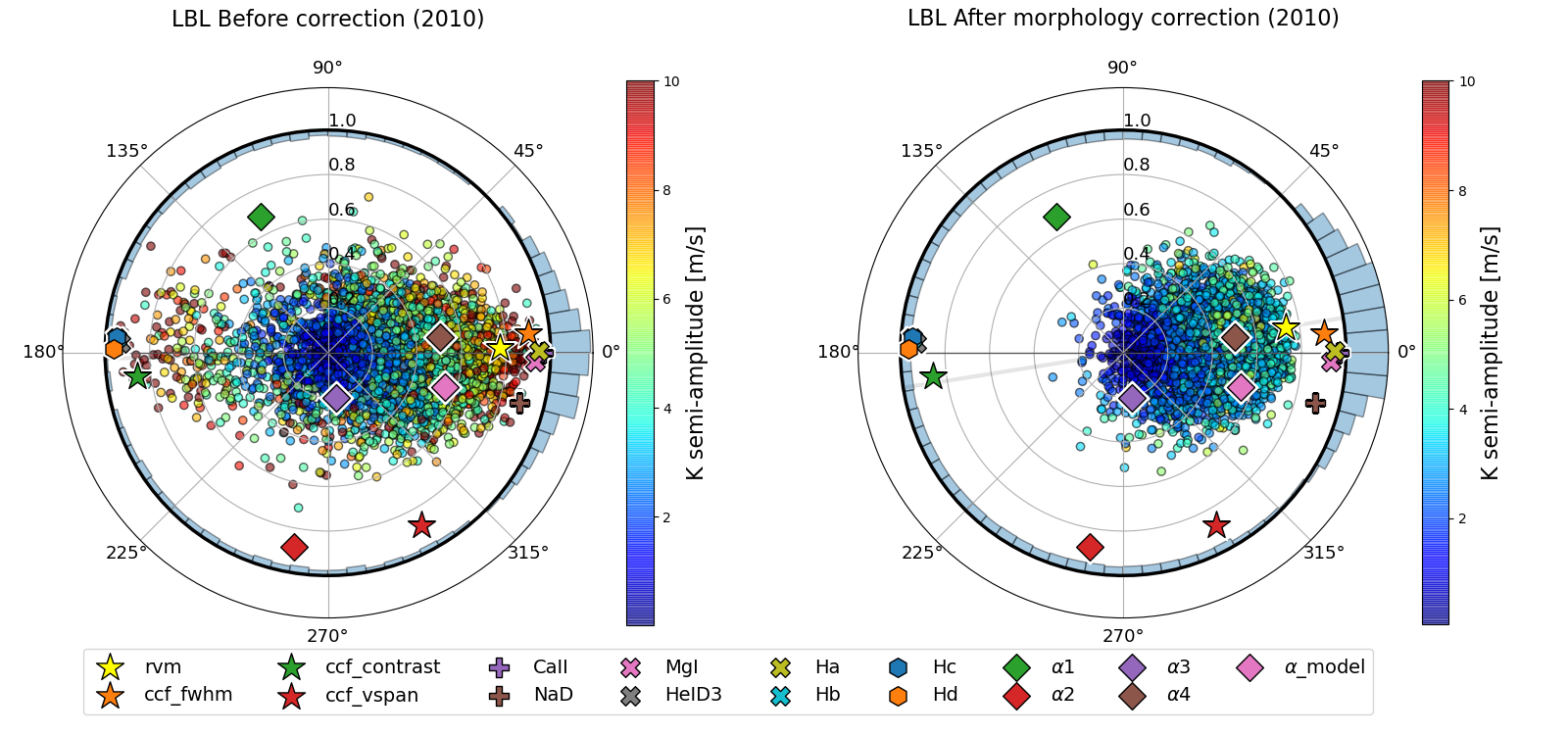}
        \caption{Polar periodogram for the 2010 HD128621 dataset obtained for a fixed period of 36 days, known to be the stellar rotation period of HD128621. Each dot represents a spectral line. Peculiar activity proxies and $\alpha_j(t)$ coefficient time series are shown as markers of different types (see legend). The radial coordinate is the weighted $\mathcal{R}$ Pearson coefficient value obtained between each LBL RV (or an activity proxy) and the best-fitting sinusoidal to the corresponding LBL RV (similar to the GLS periodogram power). The polar angle is the phase of the fitted sinusoidal, the reference $\phi=0^\circ$ is defined as the phase obtained for the CaII\,H\&K time series. The external histogram indicate the angular distribution compared to an isotropic distribution highlighted by the thick black circle. Histogram bins touching the most external circle are at 5 sigmas. 
        \textbf{Left: } Prior to the YARARA correction. A number of stellar lines contain anti-correlated lines ($\phi=180^\circ$) with respect to the CaII\,H\&K time series. \textbf{Right: } Same as left after YARARA post-processing. All the previous anti-correlated lines disappear and we are only left with correlated lines ($\phi=0^\circ$). }
        \label{FigPolar36days}
        
        \centering
        \includegraphics[width=18cm]{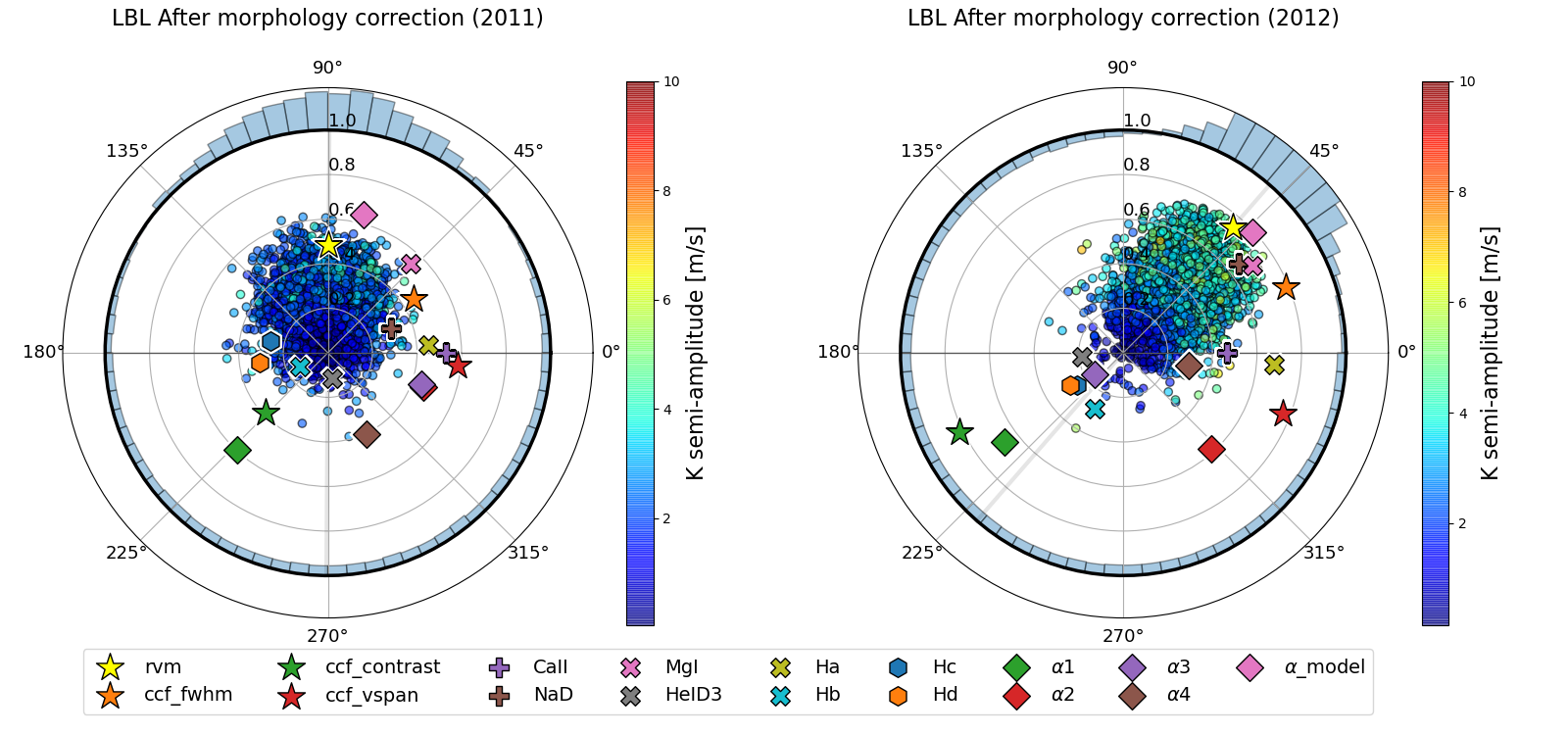}
        \caption{Polar periodogram for the 2011 and 2012 HD128621 datasets after YARARA correction. For 2012, the highest peaks in the periodogram correspond to a period of 40 days, so we display the polar periodogram at 40 days rather than 36 days. A clear time-lag between the RV variation of each line and the CaII\,H\&K (purple cross) of $\phi\sim 90^\circ$ ($9$ days) and $\phi\sim 45^\circ$ ($5$ days) is visible in 2011 and 2012, respectively. The RVs are however in phase with the multi-linear model of $\alpha_j$ (pink square) as well as with the CCF FWHM (orange star).}
        \label{FigPolar36days2011}
\end{figure*} 

To measure such collinearity effects under the assumption of a unique planetary signal in circular orbit, which allow for a fast computational time, we can simply simulate a sinusoidal variation with a given period and phase and with an arbitrary amplitude, $K$, evaluated on the time sampling of the observations; we then project this signal on the $\alpha_j$ basis with a linear least-square and measure the residual amplitude, $K_{res}$. We note that because the contamination model consist of a simple multi-linear regression (see Eq.~\ref{eq:5}), the amplitude of the simulated planets does not influence the relative absorption of the signal. We simulated 18 different phases and 15'000 different periods equidistant in frequency from three days up to the length of the observational baseline. The relative absorption and standard deviation of a sinusoidal signal, after averaging the different phases, are displayed in Fig.~\ref{FigAbs} for HD10700. We observe that for the data of this star, none of the injected signals are absorbed by more than 13\% in average, which is in line with the small number of components (only four) with respect to the 406 data points fitted.
As a general rule, an increase towards longest periods is observed, simply due to the fact that long-terms signal are more degenerate than shorter ones for dense observational sampling due to the finite length of the data set.
\begin{figure*}[tp]
        \centering
        \includegraphics[width=18cm]{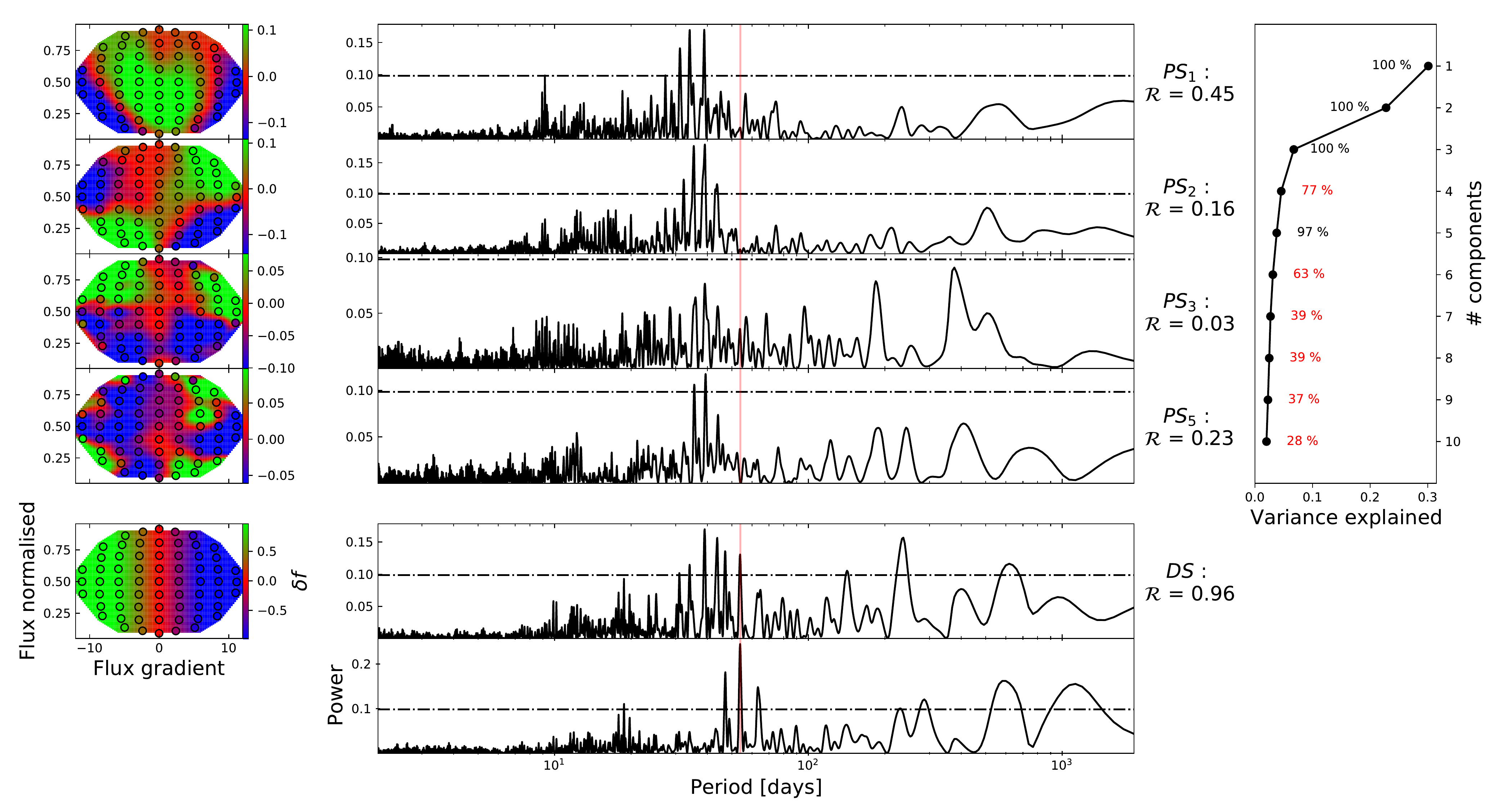}
        \caption{Shell decomposition of HD128621 spectra. The first shell is the one presenting the highest correlation with CCF RVs. The second shell is similar to a change of the line bisector and the time series of the related coefficient strongly correlated with the VSPAN of the CCF. Third shell, present mostly power at one-year and a half-year and is more related to instrumental systematics. The fourth component (fifth one according to the explained variance) seems once again to be related to stellar activity. The next components all possess a score lower than 80\% and are therefore considered as less significant. After the decorrelation by the shell coefficients $\alpha_j(t)$, the planetary injected period at 54 days (vertical red line) becomes the dominant signal, with the other peaks at 47 and 63 days being the one-year aliases.}
        \label{FigShells1}
\end{figure*} 

\begin{figure*}[tp]
        \centering
        \includegraphics[width=18cm]{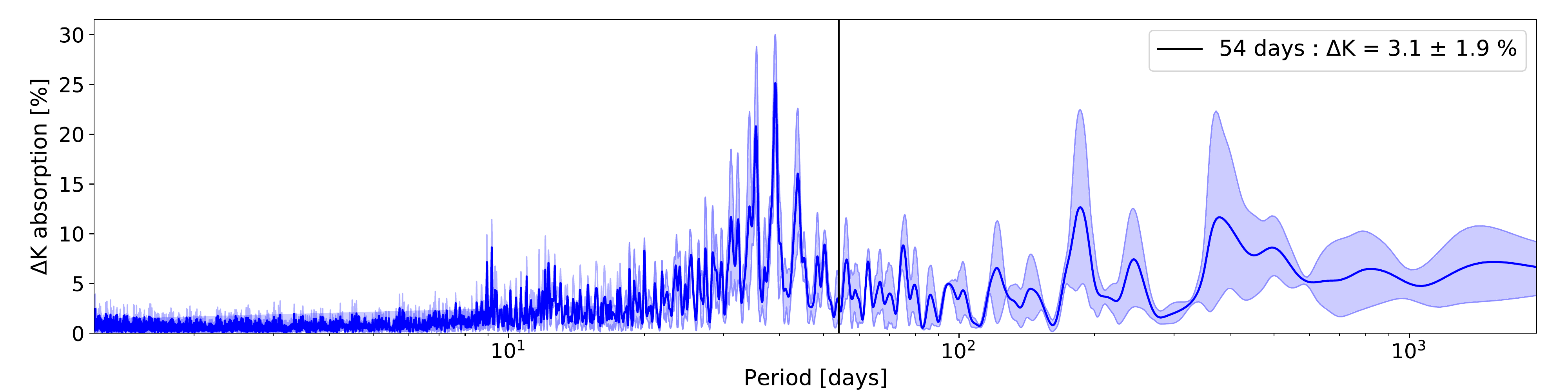}
        \caption{Measurement of the collinearity between planets in circular orbits and the $\alpha_j(t)$ basis for HD128621. A maximum of absorption of $25\pm5$\% is observed for the rotational period of the star at 39 days. It demonstrates than even if the rotational signal is strongly mitigated, a real circular planetary signal at 39-days would not have been more absorbed than this. For the injected planetary signal at 54 days, the absorption is $3\pm2$ \%. Absorption at one-year and a half-year are also noticeable.}
        \label{FigAbs2}
\end{figure*} 

For the 37-day injected planetary signal, the absorption is negligible (only 2\% in average) whereas that for the 122-day signal, a relative amplitude difference of $7\pm4$ \% is observed. Such an analysis is only valid in the case of multi-planetary non-eccentric signals, as long as there is no cross-talk between the planetary signals, which can happen in the case of poor sampling with regard to the signals. Even if the best strategy remains to included the $\alpha_j(t)$ basis coefficients inside the fitted Keplerian model, an analysis such as the one presented above, can be used to estimate with a data-driven approach by how much planetary signals can be absorbed.
This planetary signal injection-recovery test confirms that the shell decomposition is perfectly able to disentangle Doppler shift from others contamination (as instrumental ones here) and that the planetary signals are not significantly absorbed.






\subsection{Correction of stellar activity on HD128621 \label{sec:activity}}

We then tested our shell framework on HD128621 ($\alpha\,Cen\,B$, K1V, magnitude $m_v=-1.3$), a star that exhibits a similar activity  to the Sun and is thus more active than HD10700. From 2008 to 2012, the star was on the increasing part of its magnetic cycle, such that the 2010, 2011, and 2012 seasons clearly exhibit strong signatures of stellar activity at the 36-day rotational period \citep[e.g.][]{Dumusque(2012b)}. We analysed the 2008 to 2012 HARPS dataset of HD128621 post-processed by YARARA \citep{Cretignier(2021)} due to its exquisite quality and the presence of clear stellar activity features.

We first investigated the impact of the YARARA post-processing on the LBL RVs. In \citet{Cretignier(2020a)}, for the 2010 dataset, we demonstrated that most of the lines presenting strong RV signals correlated with activity were simply blended lines changing either in depth or width. This effect  produced an unexpected physical result from the CBI point of view, namely, the production of lines anti-correlated with stellar activity. Since YARARA decorrelates the flux time series at each wavelength by fitting the contrast and the full width half maximum (FWHM) of the CCF, YARARA should correct for this effect, which is indeed observed in Fig.~\ref{FigPolar36days}. In that figure, we display the polar periodogram \citep{Cretignier(2021)} at 36 days for the LBL RVs of the 2010 dataset. We also indicate several activity proxies used commonly in the literature as well as the new $\alpha_j$ shell basis coefficients. We clearly see that lines anti-correlated with the S-index at $\phi=180^\circ$, as by definition $\phi=0^\circ$ corresponds to the phase of the S-index time series, completely disappear after YARARA post-processing. We are thus left with lines that are all correlated to some level to the S-index, which is coherent with the CBI theory.



In Figs.~\ref{FigPolar36days} and \ref{FigPolar36days2011}, we can observe that for each active season (2010 to 2012) the distribution of spectral lines in the polar periodogram is clearly anisotropic in the angular coordinate, which indicates an excess of power at the rotational period in a GLS periodogram. We note that the highest peak in the periodogram was found around 40 days rather than 36 days in 2012. Also, different phase lags can be observed between the average phase of all the spectral lines and the phase of the S-index. When it is in phase in 2010, the phase lag in 2011 and 2012 is $\sim$90$^\circ$ and $\sim$45$^\circ$, respectively. We note that the precise phase lag measurement depends on the choice of the period due to the covariance between both parameters. Such phase lags explains why linearly decorrelating the RV time series with the S-index time series does not mitigate stellar activity and why kernel regression \citep[e.g.][]{Lanza:2018aa} or Gaussian Process (GP) regression \citep[e.g.][]{Rajpaul(2015)} allow for an improved  mitigation stellar activity signals. 

\begin{figure}[tp]
        \centering
        \includegraphics[width=9cm]{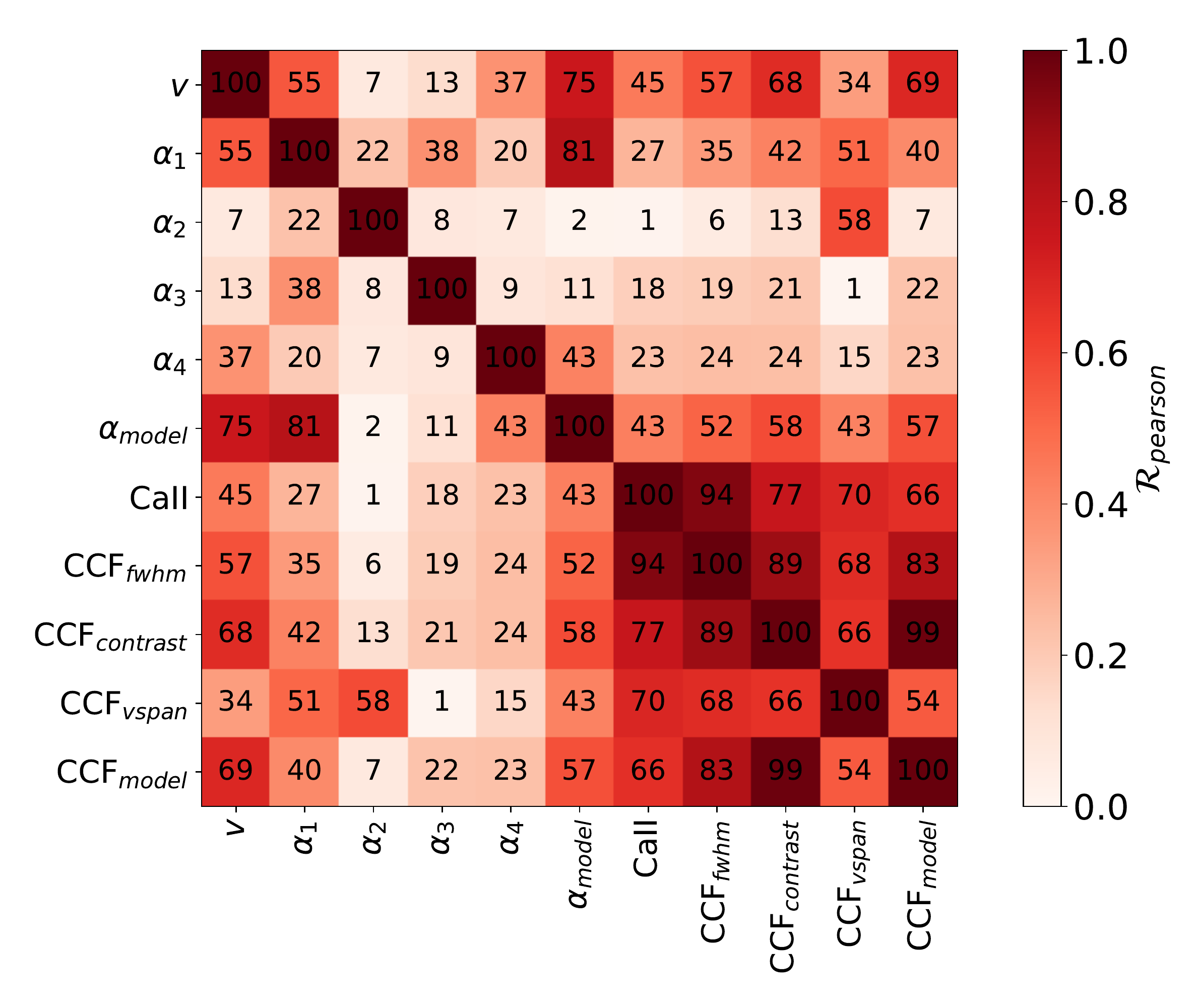}
        \caption{Correlation matrix of the CCF RVs time series, $v(t)$, the shell basis coefficients, $\alpha_j(t),$ and the full model fitted with other common activity proxies. The different shells basis coefficients show weak correlations with the CCF moments highlighting an indirect connection between shells and CCFs.}
        \label{FigCorrelation}
\end{figure} 

\begin{figure}[tp]
        \centering
        \includegraphics[width=9cm]{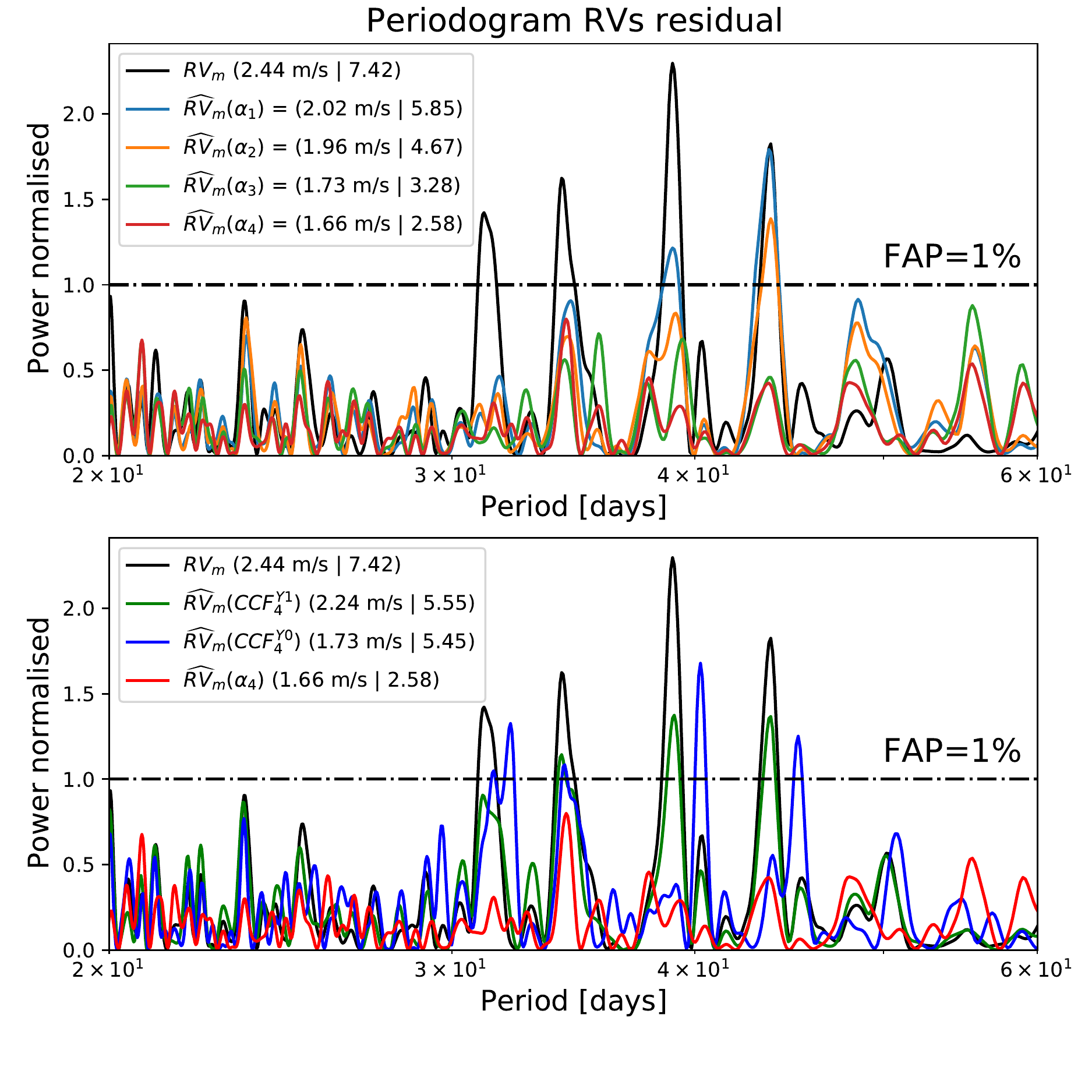}
        \caption{Periodograms of the residuals RVs parametrised by the model complexity (name convention as in Eq.~\ref{eq:5}). All the periodograms are normalised according to their respective 1\% FAP level (horizontal dotted-dashed line). The RV rms of the residuals for each model is indicated in the legend as well as the integrated power between 32 and 45 days. \textbf{Top:} Results obtained by increasingly adding shell basis coefficients into the multi-linear model of Eq.~\ref{eq:5}. Once three shell basis coefficients are fitted, the rotational period is no more significant (green and red). \textbf{Bottom:} Comparison between the residuals obtained after fitting for four shell basis coefficients (red) or the three classical CCF moments (FWHM, VSPAN and contrast) plus the $CaII\,H\&K$. Both CCF moments before ($CCF^{Y0}$, blue) and after YARARA processing ($CCF^{Y1}$, green) were used.}
        \label{FigComplexity}
\end{figure} 

\subsubsection{Shells decomposition of HD128621 \label{sec:activity1}}

\begin{figure*}[h]
        \centering
        \includegraphics[width=18cm]{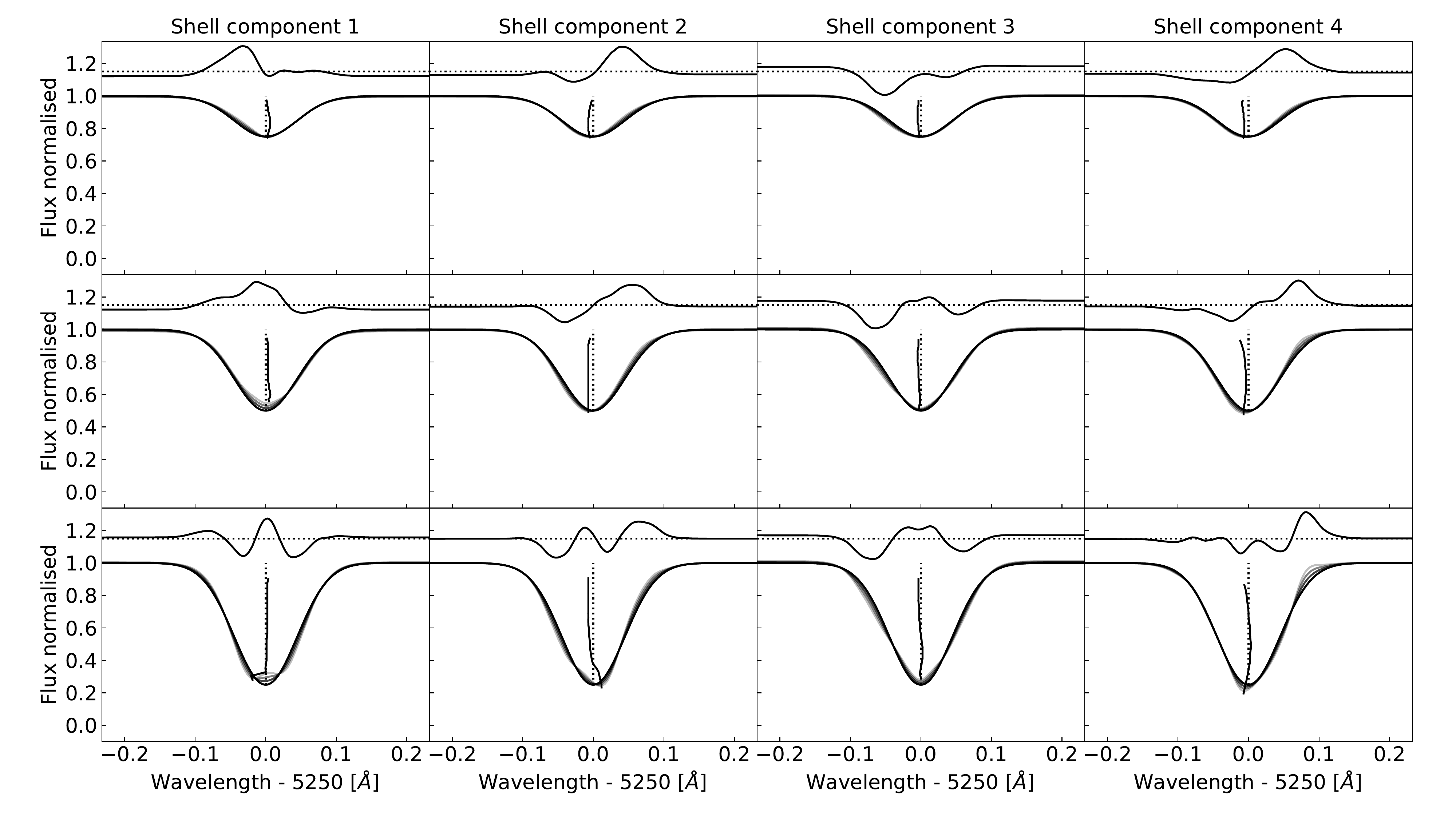}
        \caption{Inverse projection of the four significant PCA shell basis components, shown on the left of Fig.~\ref{FigShells1}, used to correct the RV time series of $\alpha$ Cen B. The inverse projection was realised for Gaussian line profiles at 5250 $\AA$ with a FWHM equivalent to 6 \kms{} and three different line depths (0.2, 0.45, and 0.7, from top to bottom). 
        The delta flux variations, $\delta f,$ evaluated along the line profile chord (see Fig.\ref{FigShell}) are indicated after magnification above the Gaussian line profile and compared with $\delta f = 0$ (horizontal dotted line). The bisector of each line profile is also indicated after magnification and compared to the initial Gaussian profile (vertical dotted lines). 
        }
        \label{FigDeprojection}
\end{figure*} 

We performed the shell decomposition described in Sect.~\ref{sec:shells} on the HD128621 YARARA post-processed data set. We display in Fig.~\ref{FigShells1} the shells fitted as well as the corresponding GLS periodogram associated with the $\alpha_j$ coefficients. As for HD10700, we decided to inject a planetary signal at 54 days and with a semi-amplitude of 1.5 \ms{} (7.5 \Mearth{} for a 0.80 \Msun{}) in order to assess the ability of shells to disentangle stellar activity from planetary signals. Only four components were found to be significant and we observe once again the remarkable smoothness of those shells, (which is not a condition required by the PCA),  indicating a smooth perturbation of the line profile induced by stellar or instrumental systematics. We clearly observe the rotational period of the star close to 36 days in all the components. However, $\alpha_3$ also present excess power at one year related to instrumental systematics, which is due to the fact that the PCA is not able to separate the different contaminations. In fact, it is expected that stellar activity and instrumental systematics will be mixed in the PCA shell basis components, even if most of the time one effect will dominate over the other. We note that the fourth basis component coefficient time series, $\alpha_5,$ is found to be significant by the cross-validation algorithm and presents power excess around the rotational period, whereas $\alpha_4$ is not found significant and is therefore discarded. For the remainder of this paper, we use the name convention from Eq.~\ref{eq:5} (indices based on the RV fit) rather than Eq.~\ref{eq:3} (indices based on PCA) to avoid gaps in components indices; therefore, we  denote the previously mentioned $\alpha_5$ as $\alpha_4$.
 
We modelled the CCF RVs -- or, equivalently, the $\alpha_{DS}(t)$ $DS$ shell basis component coefficient -- using our multi-linear regression composed of the $\alpha_j$ shell basis coefficient time series (see Eq.~\ref{eq:5}). The highest peak recovered in the RV residuals is the injected planet at 54 days, with the other peaks at 47 and 63 days standing as the one-year aliases of the injected planetary signal. This demonstrates the ability of such method to disentangle pure Doppler shift signals from stellar and instrumental contaminations. In addition, the $K$ semi-amplitude of the planet fitted on the corrected RVs is found to be 1.37 \ms{}, which is only 8\% smaller than the injected value. By performing the same simple sinusoidal signal projection as for HD10700 (see Sec.~\ref{sec:planets_injections}), a value of $3 \pm 2$\% of absorption is found (see Fig.\ref{FigAbs2}), which is compatible at $2.5\sigma$ with the measured absorption of 8\%. The slightly higher observed absorption can be explained by the interaction with pre-existing signals in the original data set. When fitting for the 54-day Keplerian signal on the planetary-free data set (scanning slightly around the period), 10 cm/s K semi-amplitude signals are typically measured, which amount to about 7\% of the injected semi-amplitude 1.5 m/s. Therefore, an uncertainty of 7\% on the recovered amplitude is expected simply from the pre-existing structures inside the initial data. The maximum absorption is about $25\pm5$\% at 39 days, which matches the suppression of the 39-day signal in the GLS periodogram. Therefore, even a real planet at 39 days would not be  absorbed at a higher level than 25\%, on average. This can be explained by the fact that our shell framework is optimised to model the change of phase and period of the stellar activity signal seen on HD128621, and thus cannot properly model a pure sinusoidal variation induced by a circular planet. This is also an advantage brought on by multi-linear models that are more rigid than kernel or GP regressions and that are therefore less likely to absorb planetary signals.



By looking at Fig.~\ref{FigShells1}, we can clearly see that the $DS$ shell basis component also contains an activity component, as shown by the excess of power around 36 days. This behaviour does not, in fact, contradict the stellar activity. Indeed, from the CBI point of view, even if a stellar line is affected differently depending on some underlying properties, such as its formation depth in the stellar atmosphere, the element, or its wavelength, we can naturally expect that all the lines are affected by a common average effect; therefore, a real Doppler shift. This has already been seen in \citet{Cretignier(2020a)}, when the authors derived a relation between inhibition of the convective blueshift and depth of spectral lines (see Fig.~8 of that paper). As soon as blended lines were removed, 90\% of the lines presented a RV signal correlated with stellar activity, and the minimal amplitude was 3.3 \ms{}.

The presence of a real Doppler shift component from stellar activity is not a substantial issue (as already explained in Sect.~\ref{sec:shells}). Indeed, even if the CBI can present a real Doppler shift component, it will also induce line distortion, which would be modelled by the different shell basis components. Thus, CBI is  not solely a pure Doppler shift. By investigating the PCA shell basis components, the first one is the most related to the inhibition of the convective blueshift, since the $\alpha_1$ coefficient time series is  strongly correlated with the CCF RVs ($\mathcal{R}= 0.45$, see correlation in Fig.~\ref{FigShells1}). 

For the forthcoming analyses, we redid the shell decomposition without the injected planet to avoid the complexity brought by this extra signal. We note that we obtained the same shell basis, as the planetary signal is completely captured by the Doppler shell basis component.

We performed a comparison between the shell basis component coefficients and classical activity proxies, namely, the CCF moments and CaII\,H\&K. The correlation matrix is displayed in Fig.~\ref{FigCorrelation}. The CCF moments are computed without the YARARA correction for stellar activity, since after this step, the CCF contrast and FWHM  no longer present any stellar activity signatures. However, since the CCF moments and CaII\,H\&K are affected by an extended magnetic trend, we systematically fit a second order polynomial function on all the time series to produce an homogeneous analysis that is focussed on the rotational period. As we can see, when the shell decomposition is performed without the injected planet, the correlation between $\alpha_1$ and the RV time series is even stronger ($\mathcal{R}= 0.55$ while it was 0.45 in Fig.~\ref{FigShells1}). The second shell basis component also captures stellar activity as it is strongly correlated with the VSPAN of the CCF ($\mathcal{R}=0.58$) and shows a phase lag of 90$^\circ$ with the RVs, which explains the low correlation among them ($\mathcal{R}= 0.07$). Overall, we observe that some correlations exist between the CCFs moments and the shell basis component coefficients, $\alpha_j$, which clearly indicates an indirect connection between the CCF and shell spaces. 


\begin{figure*}[h]
        \centering
        \includegraphics[width=18cm]{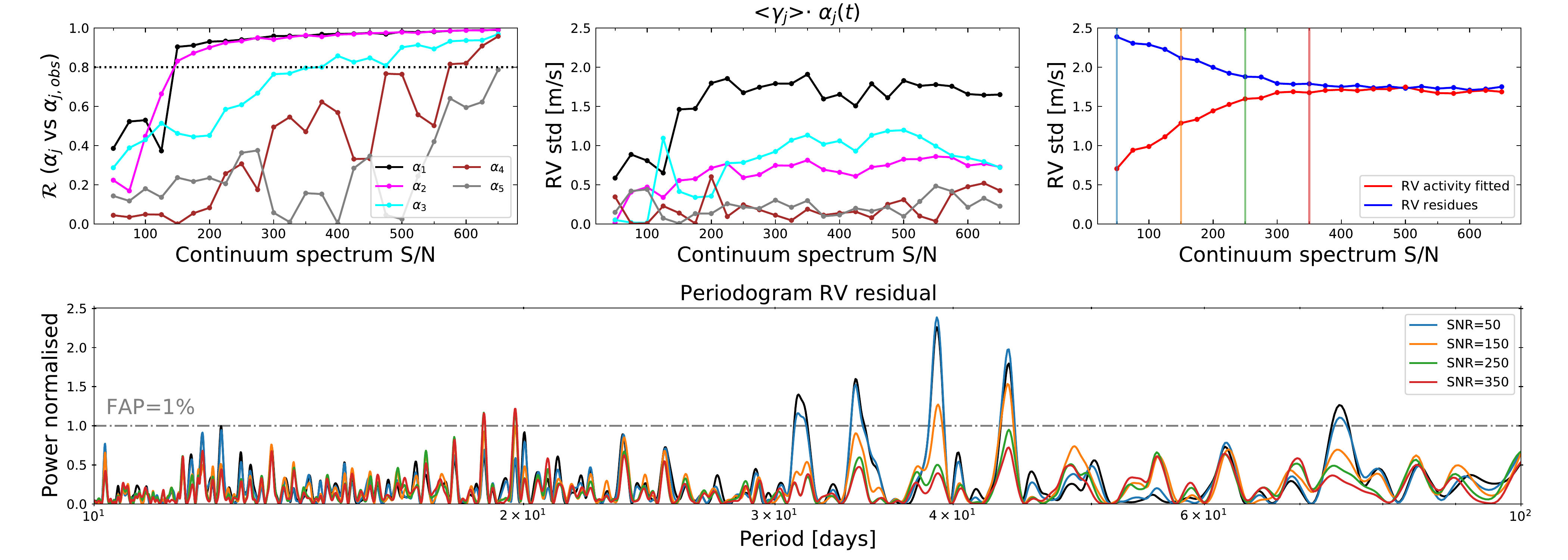}
        \caption{Noise injection tests performed on HD128621 to evaluate the performance of our shell framework in mitigating stellar activity depending on the continuum signal-to-noise ratio, \snr{}. The cross-validation algorithm to select only shell basis components with significant variance was disabled and five components were used for the fitted model. \textbf{Top left: } Correlation between the noise-affected shell basis coefficients $\alpha_j$ and their respective coefficients without extra-noise injected $\alpha_{j,obs}$. A shell basis component can be considered as recovered once its related coefficient $\alpha_j$ exceeds a correlation value with $\alpha_{j,obs}$ higher than $\mathcal{R}=0.80$ (dotted line). This threshold was selected as the level that the shell basis component $\alpha_5$, which is associated with noise in the observations, never reaches. \textbf{Top middle: } RV rms of the projected $\alpha_j(t)$ on the raw RV time series. \textbf{Top right: } RV rms of the residual RVs linearly decorrelated from the shell basis coefficients $\alpha_j$ (blue), as well as RV rms of the fitted RV activity model (red). The result for four different interesting \snr{} values (vertical lines): 50, 150, 250 and 350 are discussed in detail in the text. \textbf{Bottom: } Periodograms of the residual RVs obtained after our shell framework correction for the four selected \snr{} levels. The periodogram of the uncorrected RVs are displayed in black as a comparison. The power of each periodogram was normalised by the 1\% FAP level.}
        \label{FigNoise}
\end{figure*} 

We note that no shell basis component alone is able to correct efficiently for the stellar activity in RV. It is only when combining all of them together that the stellar activity mitigation is optimal. This is visible in Fig.~\ref{FigPolar36days} and \ref{FigPolar36days2011}, where none of the $\alpha_j$ coefficients are in phase with the RVs simultaneously for 2010, 2011, and 2012; whereas it is only the complete multi-linear model (pink square) that gives (roughly) the correct phase for each observational season. This indicates that several different perturbations of the line profile are in play, likely induced by different type of active regions (spots versus faculae) or active regions with different intrinsic properties evolving over time (temperature,  size, location, or magnetic field strength). 

In Fig.~\ref{FigComplexity}, we display the periodogram of the residuals RVs as a function of the number of $\alpha_j$ coefficients fitted. It can be seen that the rotational period is strongly mitigated after the inclusion of three shell basis coefficients into the multi-linear model. The second coefficient, $\alpha_2,$ is barely effective in improving the RVs due to the previously mentioned phase lag of $90^\circ$ with respect to the RVs. To demonstrate that shell basis coefficients and CCFs moments are different, despite their correlations, we also display in the bottom panel of the same figure the residuals obtained by fitting the contrast, FWHM, and VSPAN, as well as CaII\,H\&K using a similar multi-linear model. The obtained result clearly demonstrates that the averaging nature of the CCF produces moments that are not as good as the shell basis coefficient in mitigating stellar activity. 

Interestingly enough, we observe that both model produce similar rms for the RV residuals, but completely different periodogram. Using our shell framework, the RV rms over the five years of HD128621 observations is improved from 2.44 \ms{} down to 1.73 \ms{} (against 1.72 \ms{} using CaII\,H\&K and CCF moments). This example demonstrates that RV rms is not an optimal metric in the context of stellar activity and a periodogram remains a more relevant statistical tool. For that reason, we also computed a second score metric defined as the integrated power between 32 and 45 days, which is the period range capturing the 36- and 40-day signals (rotational period measured in 2010 and 2012) and their respective one-year aliases. Despite the similar rms, the model obtained from the shells clearly demonstrate an higher ability to absorb power at the activity-related signals with a score reduced by a factor 2.51 against only 1.29 for CCF moments. We also note that CCF moments obtained on spectra corrected of YARARA do not present any more activity signatures and cannot be used to correct the RVs.

Once we have demonstrated what shells can do, we can try to better understand what they represent. Getting a feeling about the line profile distortion is not always evident from the spectrum-folded representation of the shells. In order to better perceive the meaning of the PCA shell basis components, we project them back onto the classical line profile space. To do so, we performed a cubic interpolation of the shell basis components on the restricted ($f'$,$f$) domain of a spectral lines with different depths. We remind that because of the degeneracies brought by blended lines, no bijection exist between shells and line profile and therefore such transformation should not be fully interpreted as the real line profile modification but, rather, it ought to carry an equivalent interpretation of the shells into the classical line profile space.

We generated three Gaussian lines at 5250 $\AA$ with a FWHM of 6 \kms{}\footnote{equivalent to the mean value of the HD128621 CCFs.} and three different line depths and extracted the $\delta f$ variation related to the four first PCA shell basis components displayed on the left of Fig.~\ref{FigShells1}. We show the result of the inverse projection in Fig.~\ref{FigDeprojection}. We observed that the $PS_1$ component is roughly equivalent to a bisector change for shallow lines, a core depth variation for medium lines, and a core broadening for deep lines. Interestingly enough, all those modifications are line profile variations expected from a stellar activity point of view and are here all inter-connected within the same shell basis component and related coefficient $\alpha_1(t)$. Carrying out a similar exercise based on $PS_2$, we find that this coefficient is equivalent to a bisector change, independently of the line depth. This is coherent with the strong correlation observed between $\alpha_2(t)$ and CCF VSPAN (see Fig.~\ref{FigCorrelation}). The third component $PS_3$ seems equivalent to a wings broadening, independent here again of the line depth. Finally, the last component $PS_4$ seems once again to be related to a change in the bisector.



\subsubsection{Stellar activity mitigation with respect to S/N \label{sec:noise_sim}}

In opposition to the CCFs that can be seen as the most optimal S/N projection of the spectrum, shells are at a lower S/N due the to the less compact nature of the projection. For that reason, a reasonable question to address is the required S/N of the observations needed to correct for stellar activity using our shell framework. We note that throughout the paper, we  use the S/N$_{cont}$ notation to refer to the S/N of spectra in the continuum, to avoid confusion with the S/N of shells.

To test the S/N$_{cont}$ required to correct for the activity signal seen on HD128621, we injected different level of photon-noise at the spectral level. In total, we performed 26 different realisations of S/N$_{cont}$ ranging from 50 to 650 by step of 25. More details about the precise process of noise injection can be found in Appendix~\ref{appendix:c}. Despite the exquisite quality of the HD128621 observations, it should be noted that the S/N$_{cont}$ of the observations are flat-field limited and therefore they never exceed (on average)  the full spectrum S/N$_{cont} \sim 650$ on HARPS \citep[e.g.][]{Cretignier(2020b)}

We performed the same shell decomposition as presented in the previous section, except that the cross-validation algorithm was not applied on the PCA shell basis components, as it would select fewer components at low S/N$_{cont}$ (see below), which is not the purpose of the present evaluation. For the same reason, the S/N$>1$ criterion for each point in the shell was not applied, which implies that the same shell space was used for all the simulations.  We chose to fit the five components in a constant way in order to get a similar model complexity for all the simulations, while recalling that four were found to be significant (as shown in Sect.~\ref{sec:activity}) at the nominal S/N$_{cont}$ values of the observations. The fifth component, which is therefore noise-dominated, will allow us to control the noise behaviour. The projection of the shell basis coefficients $\alpha_j$ was performed on the raw RV time series without extra-noise injected. In that way, the simulations solely measure the degradation of the shells with respect to noise.

In the top left panel of Fig.~\ref{FigNoise}, we represent the correlations between the shell basis coefficients without extra-noise
injected, $\alpha_{j,obs}$, and the shell basis coefficients, $\alpha_j$, obtained after noise injection. In the top middle panel, we show the RV rms of the shell basis coefficients $\alpha_{j,\,obs}$ projected onto the time-domain using the $\gamma_{i,j}$ averaged over all the lines, $<\gamma_{j}>$ (see Eq.~\ref{eq:5}). In the top right panel, we display the RV rms of the modelled activity, as well as the RV rms of the residuals after this model has been fitted to the RVs for different \snr{}. Finally, in the bottom panel, we show the periodogram of the RV rms of the residuals after fitting for the shell basis components obtained for four typical S/N$_{cont}$ simulations, namely, S/N$_{cont}=50$, 150, 250, and 350. In the following, we  discuss  the result for those four typical \snr{}.

At S/N$_{cont}=50$, our lowest S/N$_{cont}$ injection test, the shell basis components are mostly noise-dominated and no significant improvement is observed in the RVs after correcting them using our shell framework. At S/N$_{cont}=150$, the two first shells basis coefficients, $\alpha_1$ and $\alpha_2$, that are related to stellar activity are clearly detected ($\mathcal{R}>0.80$) but we only performed a partial correction since $\alpha_3$ is not yet recovered. At S/N$_{cont}=250$, the power of the periodogram peaks related to stellar activity is now smaller than the 1\% false alarm probability level (see bottom panel of Fig.~\ref{FigNoise}). It is only after S/N$_{cont}=350$ that the third shell basis coefficient $\alpha_3$ is recovered. From this S/N$_{cont}$ value, the RV fitted model stop to significantly mitigate stellar activity. 
From S/N$_{cont}=350$ to S/N$_{cont}=650$ the only notable observation is the recovery of $\alpha_4$, which becomes fully detected at S/N$_{cont}=550$, but we already demonstrated in the previous section that most of the stellar activity correction was performed by the fit of $\alpha_1$, $\alpha_2$, and $\alpha_3$, which is also highlighted in the top middle panel of Fig.~\ref{FigNoise}; here, we see that the $\alpha_4$ shell basis coefficient projected onto the RVs is only responsible for a RV rms smaller than 50 \cms{}. We therefore conclude that the minimal \snr{} required to optimally correct for the stellar activity signal in HD128621 is 250. 

\begin{figure}[t]
        \centering
        \includegraphics[width=9cm]{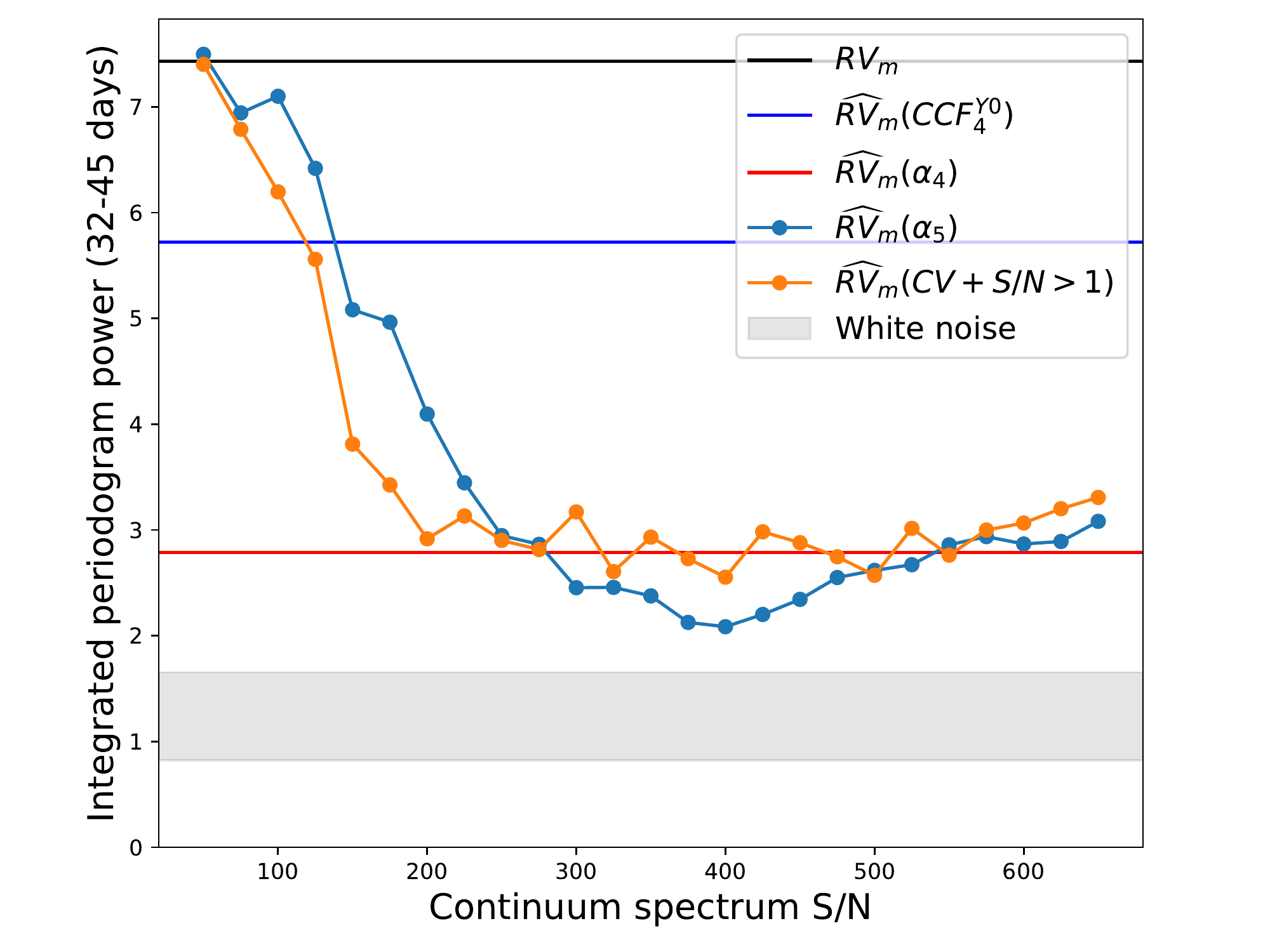}
        \caption{Noise injection tests performed on HD128621 to evaluate the performance of our shell framework in mitigating stellar activity. In this figure, we show the dependence between the integrated power of signals between 32 and 45 days, which contain mostly stellar activity signal, and the \snr{}. Previous values mentioned in Fig.~\ref{FigComplexity} and obtained at observed \snr{} values ($\sim650$) are reported (solid horizontal lines). Activating the cross-validation algorithm to select only shell basis component with significant signal and imposing the S/N$>1$ criterion to reject noisy shell elements allow to mitigate significantly stellar activity starting at a S/N$_{cont}=200$ (orange curve) instead of S/N$_{cont}=250$ (blue curve). One thousand random realisations of white noise with the time sampling of the data were produced to get a reference power value, given at $\pm 1\sigma$ (gray area). }
        \label{FigNoise2}
\end{figure}

The present limitation in \snr{} is mainly induced by noise in the shells that are fitted by our model.
We perform a second iteration of the previous simulations, this time including  the S/N$>1$ criterion on the shell elements and the cross validation algorithm. We display, for both cases in Fig.~\ref{FigNoise2}, with or without the S/N$>1$ criterion and cross-validation, the integrated power between 32 and 45 days of the residuals RVs, which contain most of the stellar activity signal. The number of significant components found by the cross-validation algorithm is 0, 2, 3, and 4 for S/N$_{cont}=50$, 150, 250, and 350, respectively. This validates the ability of our cross-validation algorithm to reject noisy components. Moreover, by imposing S/N$>1$ and the cross validation, we can strongly mitigate stellar activity starting at a S/N$_{cont}=200$ instead of S/N$_{cont}=250$ with the five-component model.

It should be noted that those noise injection tests do not take into consideration the degradation of the YARARA performance with the S/N$_{cont}$, nor the degradation of the RVs themselves -- thus, these are  optimistic results. Furthermore, those threshold S/N$_{cont}$ limits are solely valid for the present $\alpha$ Cen B activity signal. Based on real data, these tests however do demonstrate  -- similarly to what was done by \citet{Davis(2017)} on simulated data -- that a very high S/N$_{cont}$ is required to correct for stellar activity signals in RV measurements. This should certainly be considered for current and further RV surveys aiming at confirming or finding the smallest amplitude planetary signals. 

Considering that activity on HD128621 induces a RV rms of $\sim 1.72$ \ms{} (quadratic difference between 2.44 \ms{}, the raw RV rms before the shell correction, and 1.73 \ms{} after correction), and that this signal is scaling linearly with the $\delta f$ variation of the spectra (see Eq.~\ref{eq:1}), the minimum \snr{} value required to correct for a specific activity jitter can be given by:
\begin{equation}
\label{eq:snr}
    S/N_{cont,min} = \frac{250}{std(RV_{act})/1.72} = \frac{430}{std(RV_{act})}
.\end{equation}

Reaching such a high \snr{} is not achievable for any star and telescope that put strong constraints on the observational strategies needed to correct for stellar activity with the present method. We note, however, that the shell method presented here is a data-driven approach that needs to detect the signal at the spectral level (or shell level in our case) and thus it is unlikely that other data-driven approaches could do better. On a positive note, HD128621 is not a peculiar active star considering its mean \logrhk{}$=-4.94$ value and, therefore, the value of S/N$_{cont}=250$ can be considered as representative for this spectral type.

Once the rotational peaks disappear, new periods become significant around 19 days. It is expected that stellar activity produces not only power excess at the rotational period but also at the rotational harmonics \citep{Boisse(2011), Borgniet(2015),Collier(2019)}. Nevertheless, this 19-day signal could also satisfy the period of a potential transiting planet observed with Hubble \citep{Demory(2015)} to confirm the debated existence of $\alpha$ Cen B b \citep{Dumusque(2012),Hatzes(2013),Rajpaul(2016)}. However, when extracting the Keplerian solution, the transiting phase of the signal did not match the potential transit ephemeris and we therefore believe that this signal is the first harmonic of the stellar rotation. The analysis of solar data \citep{Dumusque(2021)} by our shell method will be presented in a forthcoming paper. We can, however, already mention that a similar result was found -- namely, the finding that the rotational period can be fully corrected but its first harmonic tends to be more resistant to our corrections. One solution could be to include the time derivative of the proxies in the modelling \citep{Aigrain(2012)}. We note that, intuitively, we could expect that spot and faculae, which are known to contribute in various ways to the RV budget \citep{Meunier(2015)}, could produce power at different harmonics due to the respectively even and odd nature of their RV signal according to the stellar meridian \citep{Dumusque(2014)}. The persistence of a RV signal at the first harmonic of the rotational period is perhaps a sign that our method is less sensitive to specific active regions. It is however unclear for the moment if this is due to the fact that shells cannot detect their signature or because the multi-linear model is overly simplistic.

\section{Discussion and conclusion\label{sec:conclusion}}

 In this paper, we present a novel approach to correcting for non-Dopplerian signatures in stellar spectra, such as those introduced by stellar activity and instrumental signals. The method consists of projecting stellar spectra into a dimensional space where the spectral information is averaged-out, but not as much as in the CCF space. Contrary to the CCF, a spectrum projected into this new space contains information about the line depth and chromaticity, which are known to be influenced by stellar signals.

Mathematically, the projection consists in looking a spectrum in the space ($f$, $f'$, $\delta f$), namely, the flux derivative, normalised flux, and flux difference between the spectrum and a master spectrum divided by wavelength. As $\delta f/f' = v/c$, where $v$ is the Doppler shift and $c$ the speed of light, such a space is particular suitable for the study Dopplerian and non-Dopplerian components (see Sect.~\ref{sec:folding}). A spectrum projected into this space, which we call a raw shell, is then binned in the ($f$, $f'$) space to increase the S/N and produce what we call a shell. The time series of shells can then be fitted using a PCA algorithm to produce an orthogonal shell basis made of the principal components (Sect.~\ref{sec:shells}). We then add an extra Doppler shell basis component and perform a Gram-Schmidt algorithm to have all shell basis component orthogonal to each other. The time series of the fitted shell basis coefficients, excluding the one from the Doppler shell basis, are ultimately used to linearly model non-Doplerian RV variations.

We demonstrate that the framework adopted here does not absorb any real planetary signal, but it  can effectively mitigate instrumental systematics (Sect.~\ref{sec:planets_injections}) and the stellar activity signal seen at the stellar rotational period (Sect.~\ref{sec:activity}). Planetary signal injection tests performed on HD10700 show that our shell framework allows us to correct for instrumental systematics that is found at a period of a year and half a year, while absorbing the planetary signal by not more than 13\% in the worst case. In the case of HD128621, we show that a planetary signal is not altered when applying our shell framework, and furthermore, the stellar activity signal estimated to induce a RV jitter of 1.72 \ms{} (quadratic difference between 2.44 \ms{}, the raw RV rms before the shell correction, and 1.73 \ms{} after correction) is strongly mitigated to a non-significant level. We also noted that the first harmonic of the stellar activity signal at 19 days was difficult to correct, which may indicate that more advanced correction models should be deployed to account for the rotational harmonics.

The transformation from a spectrum into a shell is not bijective and, thus, it is not possible to project  the shell basis back into the spectrum space. We thus have to correct for systematics in the time-domain -- and not in the flux-domain. This implies that collinearity between our shell basis coefficients and planetary signals can always occur, and therefore, we have to model RV systematics and planetary signals simultaneously to prevent such eventuality, as done in SCALPEL \citep[][]{Collier(2021)}. This will be formally implemented in a subsequent paper, where we will further improve the RVs to correct for the last remaining instrumental systematics.

Working in the shell space where the representation of a spectrum is at a lower S/N than the CCF has a drawback, nonetheless. Indeed, simulations performed in Sect.~\ref{sec:noise_sim} demonstrate that our shell framework is efficient at mitigating the stellar activity at the rotational period of HD128621 only for spectra with S/N$_{cont} > 250$. This conclusion strongly motivates us to pursue previous intensive observational strategies of multi-observations per nights \citep{Dumusque(2011)b}, which is a natural strategy used to boost the S/N and mitigate shortest stellar signals such as the ones induced by oscillation modes and granulation. The spectra could be stacked on several days if the rotational period of the star is long enough, but in any case, this requires stars with intensive observational strategy and observations almost every night. An intense effort already known to be paramount to detect Earth-like planets \citep{Thompson(2016),Hall(2018),Gupta(2021)}. 

Because the noise injection tests only measure the efficiency of the PCA algorithm to recover significant shell basis components despite noise, the demonstrated 
S/N$_{cont} = 250$ limit could theoretically be relaxed if the data-driven method was replaced or combined with a physically motivated approach, making the PCA algorithm unnecessary. Therefore we want to use future studies to explore  whether the same shell basis could be used for stars of similar spectral type or if those shell bases are somehow related to the atomic physics of spectral lines.

Possible ways to improve the current work would be to fit the new proxies found here, namely, the time series of the shell basis coefficient, with a more complex model than a simple multi-linear regression, such as kernel regression, for example. However, the advantage of multi-linear regressions is the ability to estimate in a fast way the degeneracy or collinearity of the fitted shell basis coefficients with circular orbits, which is computationally far more costly with kernel regression. We could also apply our shell framework to the HARPS-N solar data \citep{Dumusque(2021)} to see to which level we are able to mitigate the stellar activity of a quiet star. Looking at the Sun could also open the possibility of studying whether the shell basis coefficients are related to resolved features on the solar surface.

\begin{acknowledgements}
We thank the referee for their really helpful and constructive feedbacks. M.C. and F.P. greatly acknowledge the support provided by the Swiss National Science Foundation through grant Nr. 184618. X.D is grateful to the Branco-Weiss Fellowship for continuous support. This project has received funding from the European Research Council (ERC) under the European Union's
Horizon 2020 research and innovation program (grant agreement SCORE No.~851555). 

This publication makes use of the Data \& Analysis Center for Exoplanets (DACE), which is a facility based at the University of Geneva (CH) dedicated to extrasolar planets data visualisation, exchange and analysis. DACE is a platform of the Swiss National Centre of Competence in Research (NCCR) PlanetS, federating the Swiss expertise in Exoplanet research. The DACE platform is available at https://dace.unige.ch. This work has been carried out within the frame of the National Centre for Competence in Research “PlanetS” supported by the Swiss National Science Foundation (SNSF).
\end{acknowledgements}

\bibliographystyle{aa}
\bibliography{Master}

\begin{thebibliography}{62}
\expandafter\ifx\csname natexlab\endcsname\relax\def\natexlab#1{#1}\fi

\bibitem[{{Aigrain} {et~al.}(2012){Aigrain}, {Pont}, \&
  {Zucker}}]{Aigrain(2012)}
{Aigrain}, S., {Pont}, F., \& {Zucker}, S. 2012, \mnras, 419, 3147

\bibitem[{{Baranne} {et~al.}(1996){Baranne}, {Queloz}, {Mayor}, {Adrianzyk},
  {Knispel}, {Kohler}, {Lacroix}, {Meunier}, {Rimbaud}, \&
  {Vin}}]{Baranne(1996)}
{Baranne}, A., {Queloz}, D., {Mayor}, M., {et~al.} 1996, \aaps, 119, 373

\bibitem[{{Barnes} {et~al.}(2017){Barnes}, {Jeffers}, {Anglada-Escud{\'e}},
  {Haswell}, {Jones}, {Tuomi}, {Feng}, {Jenkins}, \& {Petit}}]{Barnes(2017)}
{Barnes}, J.~R., {Jeffers}, S.~V., {Anglada-Escud{\'e}}, G., {et~al.} 2017,
  \mnras, 466, 1733

\bibitem[{{Basri} {et~al.}(1989){Basri}, {Wilcots}, \& {Stout}}]{Basri(1989)}
{Basri}, G., {Wilcots}, E., \& {Stout}, N. 1989, \pasp, 101, 528

\bibitem[{{Boisse} {et~al.}(2011){Boisse}, {Bouchy}, {H{\'e}brard}, {Bonfils},
  {Santos}, \& {Vauclair}}]{Boisse(2011)}
{Boisse}, I., {Bouchy}, F., {H{\'e}brard}, G., {et~al.} 2011, \aap, 528, A4

\bibitem[{{Borgniet} {et~al.}(2015){Borgniet}, {Meunier}, \&
  {Lagrange}}]{Borgniet(2015)}
{Borgniet}, S., {Meunier}, N., \& {Lagrange}, A.-M. 2015, \aap, 581, A133

\bibitem[{{Bouchy} {et~al.}(2001){Bouchy}, {Pepe}, \& {Queloz}}]{Bouchy(2001)}
{Bouchy}, F., {Pepe}, F., \& {Queloz}, D. 2001, \aap, 374, 733

\bibitem[{{Brewer} {et~al.}(2020){Brewer}, {Fischer}, {Blackman}, {Cabot},
  {Davis}, {Laughlin}, {Leet}, {Ong}, {Petersburg}, {Szymkowiak}, {Zhao},
  {Henry}, \& {Llama}}]{Brewer:2020aa}
{Brewer}, J.~M., {Fischer}, D.~A., {Blackman}, R.~T., {et~al.} 2020, \aj, 160,
  67

\bibitem[{{Collier Cameron} {et~al.}(2021){Collier Cameron}, {Ford}, {Shahaf},
  {Aigrain}, {Dumusque}, {Haywood}, {Mortier}, {Phillips}, {Buchhave},
  {Cecconi}, {Cegla}, {Cosentino}, {Cr{\'e}tignier}, {Ghedina}, {Gonz{\'a}lez},
  {Latham}, {Lodi}, {L{\'o}pez-Morales}, {Micela}, {Molinari}, {Pepe},
  {Piotto}, {Poretti}, {Queloz}, {Juan}, {S{\'e}gransan}, {Sozzetti},
  {Szentgyorgyi}, {Thompson}, {Udry}, \& {Watson}}]{Collier(2021)}
{Collier Cameron}, A., {Ford}, E.~B., {Shahaf}, S., {et~al.} 2021, \mnras, 505,
  1699

\bibitem[{{Collier Cameron} {et~al.}(2019){Collier Cameron}, {Mortier},
  {Phillips}, {Dumusque}, {Haywood}, {Langellier}, {Watson}, {Cegla}, {Costes},
  {Charbonneau}, {Coffinet}, {Latham}, {Lopez-Morales}, {Malavolta},
  {Maldonado}, {Micela}, {Milbourne}, {Molinari}, {Saar}, {Thompson},
  {Buchschacher}, {Cecconi}, {Cosentino}, {Ghedina}, {Glenday}, {Gonzalez},
  {Li}, {Lodi}, {Lovis}, {Pepe}, {Poretti}, {Rice}, {Sasselov}, {Sozzetti},
  {Szentgyorgyi}, {Udry}, \& {Walsworth}}]{Collier(2019)}
{Collier Cameron}, A., {Mortier}, A., {Phillips}, D., {et~al.} 2019, \mnras,
  487, 1082

\bibitem[{{Cosentino} {et~al.}(2012){Cosentino}, {Lovis}, {Pepe}, {Collier
  Cameron}, {Latham}, {Molinari}, {Udry}, {Bezawada}, {Black}, {Born},
  {Buchschacher}, {Charbonneau}, {Figueira}, {Fleury}, {Galli}, {Gallie},
  {Gao}, {Ghedina}, {Gonzalez}, {Gonzalez}, {Guerra}, {Henry}, {Horne},
  {Hughes}, {Kelly}, {Lodi}, {Lunney}, {Maire}, {Mayor}, {Micela}, {Ordway},
  {Peacock}, {Phillips}, {Piotto}, {Pollacco}, {Queloz}, {Rice}, {Riverol},
  {Riverol}, {San Juan}, {Sasselov}, {Segransan}, {Sozzetti}, {Sosnowska},
  {Stobie}, {Szentgyorgyi}, {Vick}, \& {Weber}}]{Cosentino(2012)}
{Cosentino}, R., {Lovis}, C., {Pepe}, F., {et~al.} 2012, in \procspie, Vol.
  8446, Ground-based and Airborne Instrumentation for Astronomy IV, 84461V

\bibitem[{{Cretignier} {et~al.}(2020{\natexlab{a}}){Cretignier}, {Dumusque},
  {Allart}, {Pepe}, \& {Lovis}}]{Cretignier(2020a)}
{Cretignier}, M., {Dumusque}, X., {Allart}, R., {Pepe}, F., \& {Lovis}, C.
  2020{\natexlab{a}}, \aap, 633, A76

\bibitem[{{Cretignier} {et~al.}(2021){Cretignier}, {Dumusque}, {Hara}, \&
  {Pepe}}]{Cretignier(2021)}
{Cretignier}, M., {Dumusque}, X., {Hara}, N.~C., \& {Pepe}, F. 2021, \aap, 653,
  A43

\bibitem[{{Cretignier} {et~al.}(2020{\natexlab{b}}){Cretignier}, {Francfort},
  {Dumusque}, {Allart}, \& {Pepe}}]{Cretignier(2020b)}
{Cretignier}, M., {Francfort}, J., {Dumusque}, X., {Allart}, R., \& {Pepe}, F.
  2020{\natexlab{b}}, \aap, 640, A42

\bibitem[{{Davis} {et~al.}(2017){Davis}, {Cisewski}, {Dumusque}, {Fischer}, \&
  {Ford}}]{Davis(2017)}
{Davis}, A.~B., {Cisewski}, J., {Dumusque}, X., {Fischer}, D.~A., \& {Ford},
  E.~B. 2017, \apj, 846, 59

\bibitem[{{de Beurs} {et~al.}(2020){de Beurs}, {Vanderburg}, {Shallue},
  {Dumusque}, {Collier Cameron}, {Buchhave}, {Cosentino}, {Ghedina}, {Haywood},
  {Langellier}, {Latham}, {L{\'o}pez-Morales}, {Mayor}, {Micela}, {Milbourne},
  {Mortier}, {Molinari}, {Pepe}, {Phillips}, {Pinamonti}, {Piotto}, {Rice},
  {Sasselov}, {Sozzetti}, {Udry}, \& {Watson}}]{Beurs:2020aa}
{de Beurs}, Z.~L., {Vanderburg}, A., {Shallue}, C.~J., {et~al.} 2020, arXiv
  e-prints, arXiv:2011.00003

\bibitem[{{Demory} {et~al.}(2015){Demory}, {Ehrenreich}, {Queloz}, {Seager},
  {Gilliland}, {Chaplin}, {Proffitt}, {Gillon}, {G{\"u}nther}, {Benneke},
  {Dumusque}, {Lovis}, {Pepe}, {S{\'e}gransan}, {Triaud}, \&
  {Udry}}]{Demory(2015)}
{Demory}, B.-O., {Ehrenreich}, D., {Queloz}, D., {et~al.} 2015, \mnras, 450,
  2043

\bibitem[{{Donati} {et~al.}(2017){Donati}, {Yu}, {Moutou}, {Cameron}, {Malo},
  {Grankin}, {H{\'e}brard}, {Hussain}, {Vidotto}, {Alencar}, {Haywood},
  {Bouvier}, {Petit}, {Takami}, {Herczeg}, {Gregory}, {Jardine}, {Morin}, \&
  {MaTYSSE Collaboration}}]{Donati(2017)}
{Donati}, J.-F., {Yu}, L., {Moutou}, C., {et~al.} 2017, \mnras, 465, 3343

\bibitem[{{Dumusque}(2012)}]{Dumusque(2012)}
{Dumusque}, X. 2012, PhD thesis, Observatory of Geneva
  <EMAIL>x.dumusque@unige.ch</EMAIL>

\bibitem[{{Dumusque}(2018)}]{Dumusque(2018)}
{Dumusque}, X. 2018, \aap, 620, A47

\bibitem[{{Dumusque} {et~al.}(2014){Dumusque}, {Boisse}, \&
  {Santos}}]{Dumusque(2014)}
{Dumusque}, X., {Boisse}, I., \& {Santos}, N.~C. 2014, \apj, 796, 132

\bibitem[{{Dumusque} {et~al.}(2021){Dumusque}, {Cretignier}, {Sosnowska},
  {Buchschacher}, {Lovis}, {Phillips}, {Pepe}, {Alesina}, {Buchhave},
  {Burnier}, {Cecconi}, {Cegla}, {Cloutier}, {Collier Cameron}, {Cosentino},
  {Ghedina}, {Gonz{\'a}lez}, {Haywood}, {Latham}, {Lodi}, {L{\'o}pez-Morales},
  {Maldonado}, {Malavolta}, {Micela}, {Molinari}, {Mortier}, {P{\'e}rez
  Ventura}, {Pinamonti}, {Poretti}, {Rice}, {Riverol}, {Riverol}, {San Juan},
  {S{\'e}gransan}, {Sozzetti}, {Thompson}, {Udry}, \&
  {Wilson}}]{Dumusque(2021)}
{Dumusque}, X., {Cretignier}, M., {Sosnowska}, D., {et~al.} 2021, \aap, 648,
  A103

\bibitem[{{Dumusque} {et~al.}(2011{\natexlab{a}}){Dumusque}, {Lovis},
  {S{\'e}gransan}, {Mayor}, {Udry}, {Benz}, {Bouchy}, {Lo Curto}, {Mordasini},
  {Pepe}, {Queloz}, {Santos}, \& {Naef}}]{Dumusque(2011)b}
{Dumusque}, X., {Lovis}, C., {S{\'e}gransan}, D., {et~al.} 2011{\natexlab{a}},
  \aap, 535, A55

\bibitem[{{Dumusque} {et~al.}(2012){Dumusque}, {Pepe}, {Lovis},
  {S{\'e}gransan}, {Sahlmann}, {Benz}, {Bouchy}, {Mayor}, {Queloz}, {Santos},
  \& {Udry}}]{Dumusque(2012b)}
{Dumusque}, X., {Pepe}, F., {Lovis}, C., {et~al.} 2012, \nat, 491, 207

\bibitem[{{Dumusque} {et~al.}(2011{\natexlab{b}}){Dumusque}, {Udry}, {Lovis},
  {Santos}, \& {Monteiro}}]{Dumusque(2011)a}
{Dumusque}, X., {Udry}, S., {Lovis}, C., {Santos}, N.~C., \& {Monteiro},
  M.~J.~P.~F.~G. 2011{\natexlab{b}}, \aap, 525, A140

\bibitem[{{Feng} {et~al.}(2017){Feng}, {Tuomi}, {Jones}, {Barnes},
  {Anglada-Escud{\'e}}, {Vogt}, \& {Butler}}]{Feng(2017)}
{Feng}, F., {Tuomi}, M., {Jones}, H.~R.~A., {et~al.} 2017, \aj, 154, 135

\bibitem[{{Figueira} {et~al.}(2013){Figueira}, {Santos}, {Pepe}, {Lovis}, \&
  {Nardetto}}]{Figueira(2013)}
{Figueira}, P., {Santos}, N.~C., {Pepe}, F., {Lovis}, C., \& {Nardetto}, N.
  2013, \aap, 557, A93

\bibitem[{{Fischer} {et~al.}(2016){Fischer}, {Anglada-Escude}, {Arriagada},
  {Baluev}, {Bean}, {Bouchy}, {Buchhave}, {Carroll}, {Chakraborty}, {Crepp},
  {Dawson}, {Diddams}, {Dumusque}, {Eastman}, {Endl}, {Figueira}, {Ford},
  {Foreman-Mackey}, {Fournier}, {F{\H{u}}r{\'e}sz}, {Gaudi}, {Gregory},
  {Grundahl}, {Hatzes}, {H{\'e}brard}, {Herrero}, {Hogg}, {Howard}, {Johnson},
  {Jorden}, {Jurgenson}, {Latham}, {Laughlin}, {Loredo}, {Lovis}, {Mahadevan},
  {McCracken}, {Pepe}, {Perez}, {Phillips}, {Plavchan}, {Prato}, {Quirrenbach},
  {Reiners}, {Robertson}, {Santos}, {Sawyer}, {Segransan}, {Sozzetti},
  {Steinmetz}, {Szentgyorgyi}, {Udry}, {Valenti}, {Wang}, {Wittenmyer}, \&
  {Wright}}]{Fischer(2016)}
{Fischer}, D.~A., {Anglada-Escude}, G., {Arriagada}, P., {et~al.} 2016, \pasp,
  128, 066001

\bibitem[{{Gupta} {et~al.}(2021){Gupta}, {Wright}, {Robertson}, {Halverson},
  {Luhn}, {Roy}, {Mahadevan}, {Ford}, {Bender}, {Blake}, {Hearty}, {Kanodia},
  {Logsdon}, {McElwain}, {Monson}, {Ninan}, {Schwab}, {Stef{\'a}nsson}, \&
  {Terrien}}]{Gupta(2021)}
{Gupta}, A.~F., {Wright}, J.~T., {Robertson}, P., {et~al.} 2021, \aj, 161, 130

\bibitem[{{Hall} {et~al.}(2018){Hall}, {Thompson}, {Handley}, \&
  {Queloz}}]{Hall(2018)}
{Hall}, R.~D., {Thompson}, S.~J., {Handley}, W., \& {Queloz}, D. 2018, \mnras,
  479, 2968

\bibitem[{{Hatzes}(2013)}]{Hatzes(2013)}
{Hatzes}, A.~P. 2013, \apj, 770, 133

\bibitem[{{Hu{\'e}lamo} {et~al.}(2008){Hu{\'e}lamo}, {Figueira}, {Bonfils},
  {Santos}, {Pepe}, {Gillon}, {Azevedo}, {Barman}, {Fern{\'a}ndez}, {di Folco},
  {Guenther}, {Lovis}, {Melo}, {Queloz}, \& {Udry}}]{Huelamo(2008)}
{Hu{\'e}lamo}, N., {Figueira}, P., {Bonfils}, X., {et~al.} 2008, \aap, 489, L9

\bibitem[{{Jurgenson} {et~al.}(2016){Jurgenson}, {Fischer}, {McCracken},
  {Sawyer}, {Szymkowiak}, {Davis}, {Muller}, \& {Santoro}}]{Jurgenson:2016aa}
{Jurgenson}, C., {Fischer}, D., {McCracken}, T., {et~al.} 2016, in Society of
  Photo-Optical Instrumentation Engineers (SPIE) Conference Series, Vol. 9908,
  Ground-based and Airborne Instrumentation for Astronomy VI, ed. C.~J.
  {Evans}, L.~{Simard}, \& H.~{Takami}, 99086T

\bibitem[{{Lanza} {et~al.}(2018){Lanza}, {Malavolta}, {Benatti}, {Desidera},
  {Bignamini}, {Bonomo}, {Esposito}, {Figueira}, {Gratton}, {Scandariato},
  {Damasso}, {Sozzetti}, {Biazzo}, {Claudi}, {Cosentino}, {Covino}, {Maggio},
  {Masiero}, {Micela}, {Molinari}, {Pagano}, {Piotto}, {Poretti}, {Smareglia},
  {Affer}, {Boccato}, {Borsa}, {Boschin}, {Giacobbe}, {Knapic}, {Leto},
  {Maldonado}, {Mancini}, {Martinez Fiorenzano}, {Messina}, {Nascimbeni},
  {Pedani}, \& {Rainer}}]{Lanza:2018aa}
{Lanza}, A.~F., {Malavolta}, L., {Benatti}, S., {et~al.} 2018, \aap, 616, A155

\bibitem[{{Lienhard} \& {Mortier}(2021)}]{Lienhard(2021)}
{Lienhard}, F. \& {Mortier}, A. 2021, in The 20.5th Cambridge Workshop on Cool
  Stars, Stellar Systems, and the Sun (CS20.5), Cambridge Workshop on Cool
  Stars, Stellar Systems, and the Sun, 155

\bibitem[{{Lisogorskyi} {et~al.}(2021){Lisogorskyi}, {Jones}, {Feng}, {Butler},
  \& {Vogt}}]{Lisogorskyi(2021)}
{Lisogorskyi}, M., {Jones}, H.~R.~A., {Feng}, F., {Butler}, R.~P., \& {Vogt},
  S. 2021, \mnras, 500, 548

\bibitem[{{L{\"o}hner-B{\"o}ttcher} {et~al.}(2019){L{\"o}hner-B{\"o}ttcher},
  {Schmidt}, {Schlichenmaier}, {Steinmetz}, \& {Holzwarth}}]{Lohner(2019)}
{L{\"o}hner-B{\"o}ttcher}, J., {Schmidt}, W., {Schlichenmaier}, R.,
  {Steinmetz}, T., \& {Holzwarth}, R. 2019, \aap, 624, A57

\bibitem[{{Marchwinski} {et~al.}(2015){Marchwinski}, {Mahadevan}, {Robertson},
  {Ramsey}, \& {Harder}}]{Marchwinski(2015)}
{Marchwinski}, R.~C., {Mahadevan}, S., {Robertson}, P., {Ramsey}, L., \&
  {Harder}, J. 2015, \apj, 798, 63

\bibitem[{{Mayor} {et~al.}(2003){Mayor}, {Pepe}, {Queloz}, {Bouchy},
  {Rupprecht}, {Lo Curto}, {Avila}, {Benz}, {Bertaux}, {Bonfils}, {Dall},
  {Dekker}, {Delabre}, {Eckert}, {Fleury}, {Gilliotte}, {Gojak}, {Guzman},
  {Kohler}, {Lizon}, {Longinotti}, {Lovis}, {Megevand}, {Pasquini}, {Reyes},
  {Sivan}, {Sosnowska}, {Soto}, {Udry}, {van Kesteren}, {Weber}, \&
  {Weilenmann}}]{Mayor(2003)}
{Mayor}, M., {Pepe}, F., {Queloz}, D., {et~al.} 2003, The Messenger, 114, 20

\bibitem[{{Meunier} {et~al.}(2010){Meunier}, {Desort}, \&
  {Lagrange}}]{Meunier(2010)}
{Meunier}, N., {Desort}, M., \& {Lagrange}, A.-M. 2010, \aap, 512, A39

\bibitem[{{Meunier} {et~al.}(2015){Meunier}, {Lagrange}, {Borgniet}, \&
  {Rieutord}}]{Meunier(2015)}
{Meunier}, N., {Lagrange}, A.-M., {Borgniet}, S., \& {Rieutord}, M. 2015, \aap,
  583, A118

\bibitem[{{Milbourne} {et~al.}(2019){Milbourne}, {Haywood}, {Phillips}, {Saar},
  {Cegla}, {Cameron}, {Costes}, {Dumusque}, {Langellier}, {Latham},
  {Maldonado}, {Malavolta}, {Mortier}, {Palumbo}, {Thompson}, {Watson},
  {Bouchy}, {Buchschacher}, {Cecconi}, {Charbonneau}, {Cosentino}, {Ghedina},
  {Glenday}, {Gonzalez}, {Li}, {Lodi}, {L{\'o}pez-Morales}, {Lovis}, {Mayor},
  {Micela}, {Molinari}, {Pepe}, {Piotto}, {Rice}, {Sasselov}, {S{\'e}gransan},
  {Sozzetti}, {Szentgyorgyi}, {Udry}, \& {Walsworth}}]{Milbourne(2019)}
{Milbourne}, T.~W., {Haywood}, R.~D., {Phillips}, D.~F., {et~al.} 2019, \apj,
  874, 107

\bibitem[{{Milbourne} {et~al.}(2021){Milbourne}, {Phillips}, {Langellier},
  {Mortier}, {Haywood}, {Saar}, {Cegla}, {Collier Cameron}, {Dumusque},
  {Latham}, {Malavolta}, {Maldonado}, {Thompson}, {Vanderburg}, {Watson},
  {Buchhave}, {Cecconi}, {Cosentino}, {Ghedina}, {Gonzalez}, {Lodi},
  {L{\'o}pez-Morales}, {Sozzetti}, \& {Walsworth}}]{Milbourne(2021)}
{Milbourne}, T.~W., {Phillips}, D.~F., {Langellier}, N., {et~al.} 2021, \apj,
  920, 21

\bibitem[{pandas~development team(2020)}]{Reback(2020)}
pandas~development team, T. 2020, pandas-dev/pandas: Pandas

\bibitem[{{Pepe} {et~al.}(2021){Pepe}, {Cristiani}, {Rebolo}, {Santos},
  {Dekker}, {Cabral}, {Di Marcantonio}, {Figueira}, {Lo Curto}, {Lovis},
  {Mayor}, {M{\'e}gevand}, {Molaro}, {Riva}, {Zapatero Osorio}, {Amate},
  {Manescau}, {Pasquini}, {Zerbi}, {Adibekyan}, {Abreu}, {Affolter}, {Alibert},
  {Aliverti}, {Allart}, {Allende Prieto}, {{\'A}lvarez}, {Alves}, {Avila},
  {Baldini}, {Bandy}, {Barros}, {Benz}, {Bianco}, {Borsa}, {Bourrier},
  {Bouchy}, {Broeg}, {Calderone}, {Cirami}, {Coelho}, {Conconi}, {Coretti},
  {Cumani}, {Cupani}, {D'Odorico}, {Damasso}, {Deiries}, {Delabre},
  {Demangeon}, {Dumusque}, {Ehrenreich}, {Faria}, {Fragoso}, {Genolet},
  {Genoni}, {G{\'e}nova Santos}, {Gonz{\'a}lez Hern{\'a}ndez}, {Hughes},
  {Iwert}, {Kerber}, {Knudstrup}, {Landoni}, {Lavie}, {Lillo-Box}, {Lizon},
  {Maire}, {Martins}, {Mehner}, {Micela}, {Modigliani}, {Monteiro}, {Monteiro},
  {Moschetti}, {Murphy}, {Nunes}, {Oggioni}, {Oliveira}, {Oshagh}, {Pall{\'e}},
  {Pariani}, {Poretti}, {Rasilla}, {Rebord{\~a}o}, {Redaelli}, {Santana
  Tschudi}, {Santin}, {Santos}, {S{\'e}gransan}, {Schmidt}, {Segovia},
  {Sosnowska}, {Sozzetti}, {Sousa}, {Span{\`o}}, {Su{\'a}rez Mascare{\~n}o},
  {Tabernero}, {Tenegi}, {Udry}, \& {Zanutta}}]{Pepe(2021)}
{Pepe}, F., {Cristiani}, S., {Rebolo}, R., {et~al.} 2021, \aap, 645, A96

\bibitem[{{Pepe} {et~al.}(2002){Pepe}, {Mayor}, {Galland}, {Naef}, {Queloz},
  {Santos}, {Udry}, \& {Burnet}}]{Pepe(2002)}
{Pepe}, F., {Mayor}, M., {Galland}, F., {et~al.} 2002, \aap, 388, 632

\bibitem[{{Rajpaul} {et~al.}(2015){Rajpaul}, {Aigrain}, {Osborne}, {Reece}, \&
  {Roberts}}]{Rajpaul(2015)}
{Rajpaul}, V., {Aigrain}, S., {Osborne}, M.~A., {Reece}, S., \& {Roberts}, S.
  2015, \mnras, 452, 2269

\bibitem[{{Rajpaul} {et~al.}(2016){Rajpaul}, {Aigrain}, \&
  {Roberts}}]{Rajpaul(2016)}
{Rajpaul}, V., {Aigrain}, S., \& {Roberts}, S. 2016, \mnras, 456, L6

\bibitem[{{Reiners} {et~al.}(2016){Reiners}, {Mrotzek}, {Lemke}, {Hinrichs}, \&
  {Reinsch}}]{Reiners(2016)}
{Reiners}, A., {Mrotzek}, N., {Lemke}, U., {Hinrichs}, J., \& {Reinsch}, K.
  2016, \aap, 587, A65

\bibitem[{{Reiners} {et~al.}(2013){Reiners}, {Shulyak}, {Anglada-Escud{\'e}},
  {Jeffers}, {Morin}, {Zechmeister}, {Kochukhov}, \&
  {Piskunov}}]{Reiners(2013)}
{Reiners}, A., {Shulyak}, D., {Anglada-Escud{\'e}}, G., {et~al.} 2013, \aap,
  552, A103

\bibitem[{{Saar}(2009)}]{Saar(2009)}
{Saar}, S.~H. 2009, in American Institute of Physics Conference Series, Vol.
  1094, 15th Cambridge Workshop on Cool Stars, Stellar Systems, and the Sun,
  ed. E.~{Stempels}, 152--161

\bibitem[{{Saar} \& {Donahue}(1997)}]{Saar-1997b}
{Saar}, S.~H. \& {Donahue}, R.~A. 1997, \apj, 485, 319

\bibitem[{{Simola} {et~al.}(2019){Simola}, {Dumusque}, \&
  {Cisewski-Kehe}}]{Simola:2019aa}
{Simola}, U., {Dumusque}, X., \& {Cisewski-Kehe}, J. 2019, \aap, 622, A131

\bibitem[{{Su{\'a}rez Mascare{\~n}o} {et~al.}(2020){Su{\'a}rez Mascare{\~n}o},
  {Faria}, {Figueira}, {Lovis}, {Damasso}, {Gonz{\'a}lez Hern{\'a}ndez},
  {Rebolo}, {Cristiani}, {Pepe}, {Santos}, {Zapatero Osorio}, {Adibekyan},
  {Hojjatpanah}, {Sozzetti}, {Murgas}, {Abreu}, {Affolter}, {Alibert},
  {Aliverti}, {Allart}, {Allende Prieto}, {Alves}, {Amate}, {Avila}, {Baldini},
  {Bandi}, {Barros}, {Bianco}, {Benz}, {Bouchy}, {Broeng}, {Cabral},
  {Calderone}, {Cirami}, {Coelho}, {Conconi}, {Coretti}, {Cumani}, {Cupani},
  {D'Odorico}, {Deiries}, {Delabre}, {Di Marcantonio}, {Dumusque},
  {Ehrenreich}, {Fragoso}, {Genolet}, {Genoni}, {G{\'e}nova Santos}, {Hughes},
  {Iwert}, {Kerber}, {Knusdstrup}, {Landoni}, {Lavie}, {Lillo-Box}, {Lizon},
  {Lo Curto}, {Maire}, {Manescau}, {Martins}, {M{\'e}gevand}, {Mehner},
  {Micela}, {Modigliani}, {Molaro}, {Monteiro}, {Monteiro}, {Moschetti},
  {Mueller}, {Nunes}, {Oggioni}, {Oliveira}, {Pall{\'e}}, {Pariani},
  {Pasquini}, {Poretti}, {Rasilla}, {Redaelli}, {Riva}, {Santana Tschudi},
  {Santin}, {Santos}, {Segovia}, {Sosnowska}, {Sousa}, {Span{\`o}}, {Tenegi},
  {Udry}, {Zanutta}, \& {Zerbi}}]{Suarez-Mascareno:2020aa}
{Su{\'a}rez Mascare{\~n}o}, A., {Faria}, J.~P., {Figueira}, P., {et~al.} 2020,
  \aap, 639, A77

\bibitem[{{Thompson} {et~al.}(2017){Thompson}, {Watson}, {de Mooij}, \&
  {Jess}}]{Thompson(2017)}
{Thompson}, A.~P.~G., {Watson}, C.~A., {de Mooij}, E.~J.~W., \& {Jess}, D.~B.
  2017, \mnras, 468, L16

\bibitem[{{Thompson} {et~al.}(2016){Thompson}, {Queloz}, {Baraffe}, {Brake},
  {Dolgopolov}, {Fisher}, {Fleury}, {Geelhoed}, {Hall}, {Gonz{\'a}lez
  Hern{\'a}ndez}, {ter Horst}, {Kragt}, {Navarro}, {Naylor}, {Pepe},
  {Piskunov}, {Rebolo}, {Sander}, {S{\'e}gransan}, {Seneta}, {Sing}, {Snellen},
  {Snik}, {Spronck}, {Stempels}, {Sun}, {Santana Tschudi}, \&
  {Young}}]{Thompson(2016)}
{Thompson}, S.~J., {Queloz}, D., {Baraffe}, I., {et~al.} 2016, in Society of
  Photo-Optical Instrumentation Engineers (SPIE) Conference Series, Vol. 9908,
  Ground-based and Airborne Instrumentation for Astronomy VI, ed. C.~J.
  {Evans}, L.~{Simard}, \& H.~{Takami}, 99086F

\bibitem[{{VanderPlas}(2018)}]{VanderPlas(2018)}
{VanderPlas}, J.~T. 2018, \apjs, 236, 16

\bibitem[{{W}es {M}c{K}inney(2010)}]{Mckinney(2010)}
{W}es {M}c{K}inney. 2010, in {P}roceedings of the 9th {P}ython in {S}cience
  {C}onference, ed. {S}t\'efan van~der {W}alt \& {J}arrod {M}illman, 56 -- 61

\bibitem[{{Wise} {et~al.}(2018){Wise}, {Dodson-Robinson}, {Bevenour}, \&
  {Provini}}]{Wise(2018)}
{Wise}, A.~W., {Dodson-Robinson}, S.~E., {Bevenour}, K., \& {Provini}, A. 2018,
  \aj, 156, 180

\bibitem[{{Zechmeister} {et~al.}(2020){Zechmeister}, {Reiners}, {Amado},
  {Azzaro}, {Bauer}, {B{\'e}jar}, {Caballero}, {Guenther}, {Hagen}, {Jeffers},
  {Kaminski}, {K{\"u}rster}, {Launhardt}, {Montes}, {Morales}, {Quirrenbach},
  {Reffert}, {Ribas}, {Seifert}, \& {Tal-Or}}]{Zechmeister(2020)}
{Zechmeister}, M., {Reiners}, A., {Amado}, P.~J., {et~al.} 2020, {SERVAL:
  SpEctrum Radial Velocity AnaLyser}

\bibitem[{{Zechmeister} {et~al.}(2018){Zechmeister}, {Reiners}, {Amado},
  {Azzaro}, {Bauer}, {B{\'e}jar}, {Caballero}, {Guenther}, {Hagen}, {Jeffers},
  {Kaminski}, {K{\"u}rster}, {Launhardt}, {Montes}, {Morales}, {Quirrenbach},
  {Reffert}, {Ribas}, {Seifert}, {Tal-Or}, \& {Wolthoff}}]{Zechmeister(2018)}
{Zechmeister}, M., {Reiners}, A., {Amado}, P.~J., {et~al.} 2018, \aap, 609, A12

\bibitem[{{Zhu} \& {Dong}(2021)}]{Zhu(2021)}
{Zhu}, W. \& {Dong}, S. 2021, \araa, 59 [\eprint[arXiv]{2103.02127}]

\end{thebibliography}

\begin{appendix}

\section{Cross-validation algorithm}
\label{appendix:a}

The question regarding the specific number of components to select is always a major concern when fitting PCA. Indeed, selecting too few components will permit undesired contaminations, whereas selecting too many components can result in overfitting and raise the levels of white noise. In order to determine the PCA components that are relevant, a cross-validation algorithm is often used.

Our cross-validation algorithm is built as follow: considering a set of observations, $T$, we randomly select $N$ subsets $t_n$ ($N=300$) containing $p$ percent of the observations ($p=80\%$). A PCA algorithm is fitted on each subset and the ten first components are selected. We choose ten components since  that number is clearly larger than the significant number of components estimated by eye (between 2 and 5, depending on the S/N). The main idea then consists of being able to compare those $10 \times N$ components with the ten components, $P$ ,fitted on the full dataset of observations ,$T$. The only required step is the classification of the $10 \times N$ components relative to the primary sample of components, $P$. The most naive approach to do so is to classify the components by reducing the 'distance' between the $10 \times (N+1)$ vectors. The size of a group measures the occurrence rate $r$ of a component. In the case where a PCA shell basis component would be affected by a single outlier observation, an occurrence rate of $r=80\%$ is expected, therefore, any PCA shell basis component with $r < 80\%$ is irrelevant.

\begin{figure*}[t]
        \centering
        \includegraphics[width=18cm]{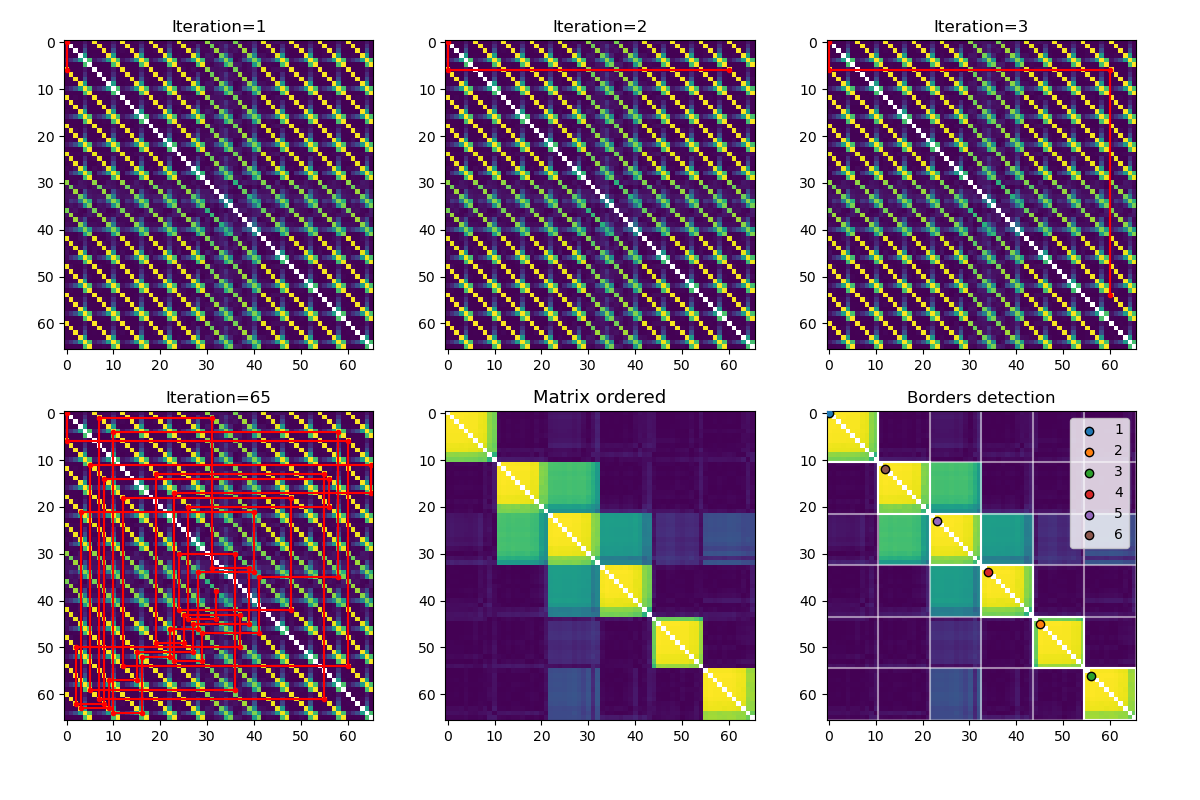}
        \caption{Illustration of the cross-validation algorithm used to measured the occurrence rate of the components through the correlation matrix $M_R$ of the vectors. The colour scale is arbitrary. Six vectors with different periods, $P,$ (from 10 to 60 days by steps of 10 days) and ten different amplitudes were used to simulate the typical output of a PCA algorithm. The $6 \cdot  (10+1) = 66$ components therefore produce a $66\times66$  correlation matrix, $M_R$. The signals at longest periods (component 4, 5 and 6) start to be collinear due to the baseline of the observations (60 days). \textbf{Top row:} First iterations of the ordering algorithm. \textbf{Bottom left:} Final chain as detected by the ranking algorithm. \textbf{Bottom middle:} Same as top left after the correlation matrix has been reordered. \textbf{Bottom right:} Border detection as given by the borders algorithm. Each cluster is clearly detected and contains a single primary components, $P,$ (colour dots). The occurrence rate, $r,$ of all the components are 100\% in that case. The borders (vertical and horizontal white lines) can be used to define a smaller matrix $R'$ (here $6 \times 6$) which can be reordered iteratively in case of bad chain ordering which did not happened in that simulation.}
        \label{FigCV}
\end{figure*}

The first step consists of selecting a metric of distance in order to classify the components into groups. A simple metric used for the distance measurement is the Pearson coefficient of correlation  $\mathcal{R}_{pearson}$, which measures the degree of collinearity between two vectors. This metric is convenient for several reasons since the value of the coefficient is bounded, swift to compute on a large dataset, and unaffected by the vectors' amplitudes. We used the pandas library \citep{Mckinney(2010),Reback(2020)} to measure the correlation matrix $M_R$ between the $10 \times (N+1)$ vectors before taking the absolute value since PCA components are free to change of sign. Hereafter, we refer to the $i^{th}$ row and $j^{th}$ column of this matrix as: $M_{R\,(i,j)}$. An example of such a correlation matrix is displayed in Fig.~\ref{FigCV}. We note that we removed the unity-diagonal because not of interest in the following analysis to measure an occurrence rate.

The measurement of the occurrence rate is divided in two steps: 
\begin{enumerate}
    \item A reordering of the rows and columns of  $M_R$
    \item A detection of sub-matrix by blocks on the reordered $M_R$
\end{enumerate}
Step 1 is accomplished in a straightforward way, and it is displayed in Fig.~\ref{FigCV}. The algorithm begins with the first element of the diagonal $M_{R\,(0,0)}$ (usually defined as the first PCA component of the primary sample $P$) and searches for the vector $M_{R\,(i,0)}$ with the highest correlation. As soon as it is located, the algorithm takes, as starting point, $M_{R\,(i,0)}$ and searches for its highest correlation, $M_{R\,(i,j)}$, and so on. If the highest correlation elements was already selected by a previous iteration, the highest correlation element not yet selected is chosen. The product of the algorithm is a $10 \times (N+1)$ vector indices describing a chain with the highest consecutive correlations between the elements. Numerically, the change in the starting point is accomplished by switching the highest correlation detection from rows to columns at each iteration (see Fig.~\ref{FigCV}). This algorithm allows a fast ranking of the matrix, and is easy to implement. The chain vector hence obtained is then used to reorder the $10 \times (N+1)$ vectors, before computing again the correlation matrix (see Fig.~\ref{FigCV}, middle bottom panel). The result of that process is the production of a matrix of correlation, where the highest correlations elements are concentrated in blocks around the diagonal. The size $s$ of those blocks (normalised by $N+1$) being the occurrence rate $r$ of the shell basis components. 

Step 2 is accomplished by an iterative algorithm to detect block edges. The matrix $M_R$ is first thresholded into a binary matrix $\tilde{M}_{R}$ such that $\tilde{M}_{R}=1$ if $M_R>\mathcal{R}_{crit}$ and zero otherwise. We begin with the first element $\tilde{M}_{R,\,(0,0)}$, and incrementally compute the sum column from $M_{R\,(0,i)}$ to $M_{R\,(i,i)}$. As soon as the sum of the $i^{th}$ column is smaller than half the column's length, we consider the right border as having been effectively reached. The index is saved and used to define the right border of the first cluster ($0$, $i$) and border left of the next cluster ($i$, $...$). The algorithm is run iteratively from $M_{R\,(i,j)}$ to $M_{R\,(j,j)}$ to form pairwise ($i$,$j$) borders limits. With those borders, several useful quantities can be obtained as the size of a block matrix, $s,$ but also the median values of the correlation inside the block matrices. The threshold $\mathcal{R}_{crit}$ can be seen as a regularisation parameter to increase or decrease the accepted 'distance' between vectors to form a cluster. 

We consider a cluster relative to a component as 'detected' if it satisfies three criteria: 1) the cluster should contains only a single primary component, $P,$ in order to avoid several components to be classified into the same cluster or clusters not related to a primary components; 2) the median of a cluster should be larger than half the largest cluster median to reject any spurious cluster formed from noise; 3) the cluster size should range between $95\%$ and $105\%$\footnote{The cluster size can exceed 100\% if components are misclassified due to spurious correlation.}, to put a strong constraint on detected cluster, where those values were found to be the best one for numerical stability after investigations. 

The algorithm of border detection solely depends on the $\mathcal{R}_{crit}$ regularisation parameter. Since this parameter is bounded and the manifold is expected to be smoothed, we simply explored several values of $\mathcal{R}_{crit}$ from 0.5 up to 0.9 by steps of 0.05. The $\mathcal{R}_{crit}$ producing the largest number of clusters detected was selected to define the borders ($i$,$j$). If several $\mathcal{R}_{crit}$ produce the same number of detected clusters, the highest $\mathcal{R}_{crit}$ is kept.

\begin{figure}[t]
        \centering
        \includegraphics[width=9cm]{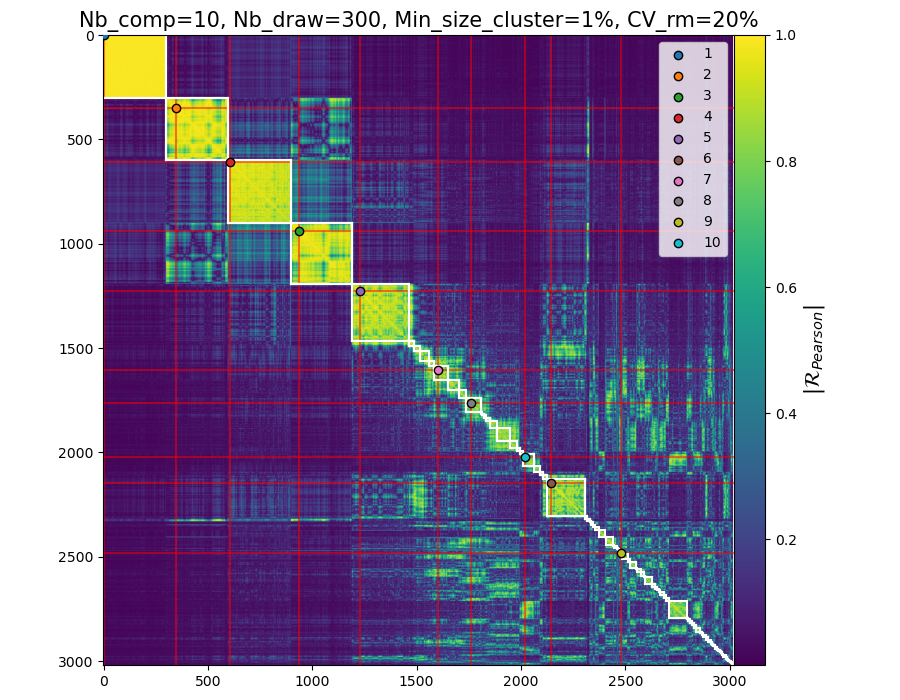}
        \caption{Illustration of the cross-validation algorithm based on the correlation matrix of the ten shells of HD10700 (vertical and horizontal red lines). Only the first four shells are detected with an occurrence rate $r>$95\%, whereas only the five first shell are not compatible with a single outlier explanation ($r>80\%$).}
        \label{FigCV2}
\end{figure}

The matrix ordering and borders detection presented above can be stuck into a wrong classification in particular if spurious correlations occur between the components, thus producing a wrong ordering of the chain. A simple way to resolve the issue consists of computing a new matrix $R'$, where the elements of the new matrix are the median inside the borders ($i$,$j$) founded previously (see bottom right panel Fig.~\ref{FigCV}). The matrix $R'$ can then be reordered according to step 1 and define the permutation to operate on the blocks of the $M_R$ matrix. After a few iterations, the clusters are well identified as visible on Fig.~\ref{FigCV2} producing reliable occurrence rate, $r,$ for the PCA components. Usually, less than four iterations are sufficient but nine are performed here for assurance. In the paper, only the n-first PCA components with an occurrence rate $r$ larger than 95\% are considered to be significant. 

\section{Impact of the master spectrum on the shell decomposition
\label{appendix:b}}

As for YARARA \citep{Cretignier(2021)}, the shell decomposition is motivated by tracking the flux variations with respect to a master spectrum assumed at a null velocity in the stellar rest-frame. When hundreds of observations are available for a star and stacked together, it is reasonable to believe that the flux variations induced by instrumental systematics, mostly fixed in the BERV rest-frame, by stellar signals, semi-periodic due to active regions evolving on the stellar surface, and by small planetary signals should be strongly averaged-out. We note that in the case of a real Doppler shift, stacking spectra without RV centred them will lead to line broadening and shallowing, a variation similar to decreasing the instrumental resolution of the master spectrum. In order to avoid that issue, spectra are first corrected from the CCF RV shift measured before being stacked together. But because those CCF RVs values may not only contains real Doppler shifts, a small error is propagated into the master spectrum. This error can come from photon noise, uncorrected instrumental systematics, stellar activity, or small planetary signals. In this section, we investigate the impact of small RV shifts.


We recall that any stationary RV signal with an amplitude larger than 5 \ms{} is pre-fitted by a Keplerian solution and removed by YARARA, since such signals can easily be recovered on an ultra stable echelle spectrograph like HARPS. Moreover, for stars intensively observed and bright which are the main targets of our work, photon-noise and instrumental systematics are usually below 1 \ms{} on HARPS, as demonstrated in \citet{Cretignier(2021)}. Finally, for slow rotating-stars with \vsini{} $\leq$ 3 \kms{}, the RV amplitude measured by SOAP 2.0 simulations are smaller than 5 \ms{} for reasonable size of active regions \citep[][]{Dumusque(2014)}. For that reason, the RV error affecting the spectra that will be used to create the master spectrum will usually not exceed 5 \ms{}, a value confirmed on $\alpha$ Cen B \citep{Dumusque(2018)} and on the Sun \citep{Dumusque(2021)}. 

Because residual RV shifts will usually not exceed a few \ms{}, a well suited data set to evaluate the impact of residual RV shifts on the master spectrum and on our shell framework is the one of HD10700 with the 3 \ms{} and 2 \ms{} planetary signals injected (see Sect.~\ref{sec:planets_injections}). Thus, we perform  our master spectrum construction and our shell decomposition once again on this dataset, except that this time we do not RV-shifted the spectra to cancel the measured CCF RVs. The obtained master spectrum will therefore be affected by the small RV residual shifts. Performing the shell decomposition, we find that this imperfect master spectrum has no impact on the algorithm. The same shells and the same RV residuals are obtained. This can be understood by the small errors introduced on the master spectrum, at most a few \ms{} compared to the average FWHM of a spectral line on HARPS, $\sim$6 \kms{}, are three orders of magnitude greater. 

\section{Noise injection implementation\label{appendix:c}}

Since photon noise follows a Poissonian distribution, the related uncertainty, $\sigma_{\gamma}$, is given as the square root of the flux $F$. The S/N$_{cont}$ of a spectrum is therefore given by $F/\sqrt{F} = \sqrt{F} = \sigma_{\gamma}$. In particular, YARARA gives output spectra that are continuum-normalised, which means that the spectra have been divided by their flux in the continuum. In this case, the photon noise in a normalised spectrum is given by $\sigma_{obs} = \sigma_{\gamma}/F = \sqrt{F}/F = (S/N_{cont})^{-1}$. In a single-observation normalised spectrum, the uncertainties in flux are dominated by the photon noise and thus the \snr{} of a continuum normalised spectrum is simply the inverse of the flux uncertainty in the continuum. This flux uncertainty can either be estimated by the error bars or by measuring 1.48 times the MAD\footnote{We note that we prefer to use the MAD rather than the standard deviation, as the MAD is less sensitive to outliers. In addition, if we multiply the MAD by 1.48, we arrive to the same value as the standard deviation in the case of data with no outliers.} in YARARA river diagram maps, considering that the river diagram maps are free of all systematics.

We note that in the case of YARARA, each spectrum is obtained by binning all the available spectra within a night. For a star with usual moderate \snr{} values and a few observations per night, the noise in the continuum of a nightly-binned YARARA spectrum is photon-noise dominated, and thus, if all spectra within a night have the same \snr{}, the \snr{} of the nightly binned spectrum should be in theory S/N$_{cont} \times \sqrt{N}$. This theoretical value should be equal to the inverse of 1.48 times the MAD measured in YARARA river diagram maps. In the case of very bright targets, such as HD128621, where each individual observation is at S/N$_{cont} \geq 400$ and dozens of observations are taken within a night, the noise in a nightly-binned YARARA spectrum is flat-field limited; this is because all the spectra within a night are reduced using the same flat field, which on HARPS has a S/N$_{cont}\sim650$. Considering 20 spectra of HD128621 at S/N$_{cont} = 400$, we should arrive at a theoretical \snr{} of $\sim 1800$ for the nightly-binned spectrum if we only consider photon-noise. However, when taking the inverse of 1.48 times the MAD in the continuum of the binned spectrum, we only reach a \snr{} value of $\sim 600$, compatible with the flat-field \snr{}, thus demonstrating that the noise in HD128621 YARARA spectra are flat-field dominated.

\begin{figure}[t]
        \centering
        \includegraphics[width=9cm]{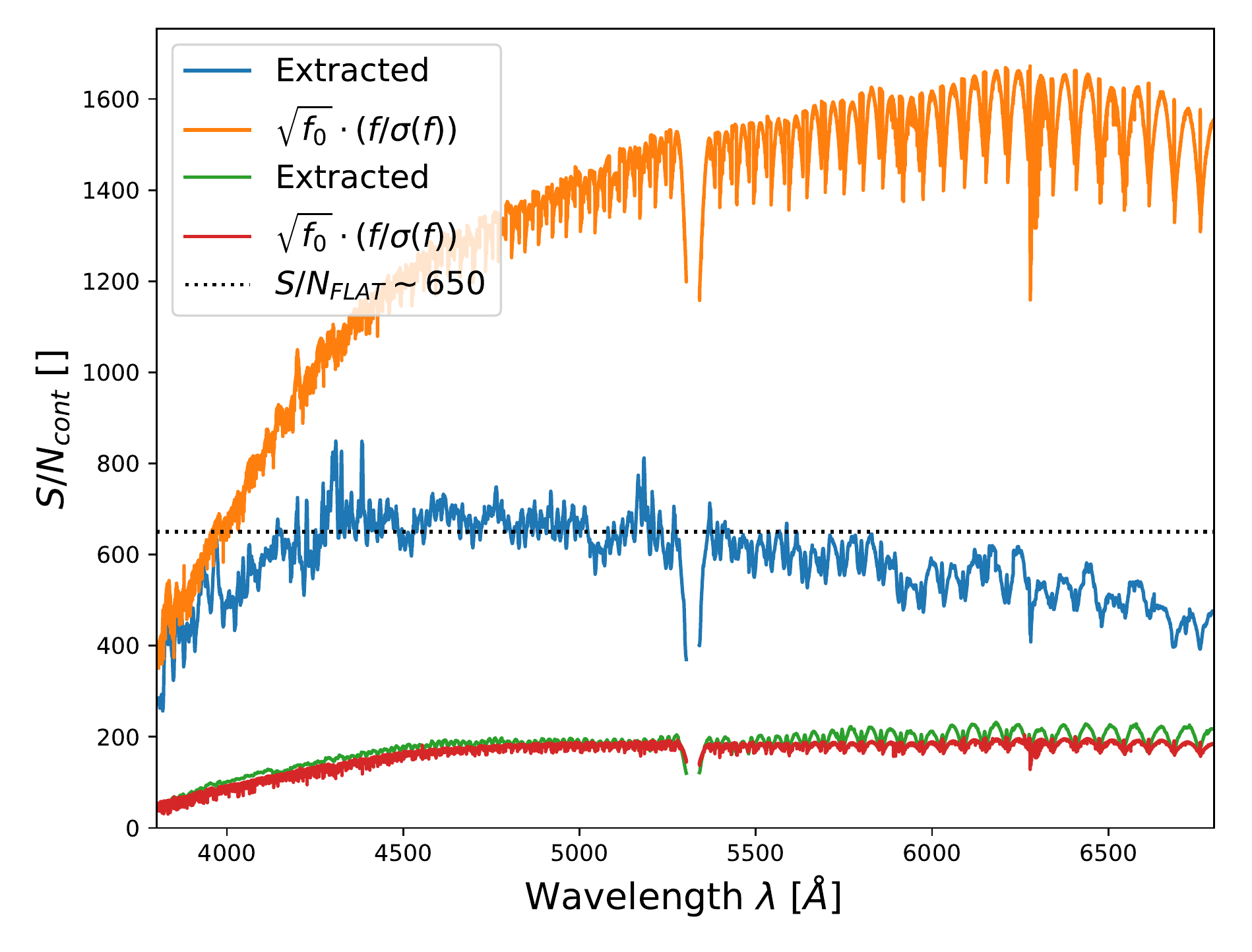}
        \caption{Examples of computed S/N$_{cont}^{obs}(\lambda)$ using the inverse of 1.48 times a rolling MAD in YARARA river diagram maps (labeled as 'extracted') and compared with the theoretical \snr{} value if only photon-noise limited. In the case of a star photon-noise limited with S/N$_{cont} \sim 200$, both methods produce similar values (green and red curves). For a spectrum flat-field limited as for HD128621 observations (blue and orange curves), the \snr{} is saturated to the \snr{} of the flat-field raw frames (black dashed line).}
        \label{FigNoiseEx2}
\end{figure}

To determine the photon noise of a nightly-binned YARARA spectrum as a function of wavelength $\sigma_{obs}(\lambda)$, or equivalently S/N$_{cont}^{obs}(\lambda)$, we assume that the noise in the continuum is a smooth function of the wavelength. Indeed, even though the \snr{} of an observed spectrum naturally change from blue to red due to several factors, as for example: i) the stellar spectral energy distribution; ii) the atmospheric extinction; iii) the use of optical fibres; and iv) the CCD response, all those effects only introduce a slow chromatic variation. We therefore consider that $1.48\times$ MAD of the YARARA river diagram maps in a window\footnote{This window has been defined as the optimal choice to detect intra-order variations as the \snr{} is always larger at the center of an order compared to the borders due to the blaze response of the echelle grating} of $\pm$ 2.5 \ang{} around $\lambda_0$ as a direct measurement of $\sigma_{obs}(\lambda_0)$. The chromatic \snr{} value of an observation S/N$_{cont}^{obs}(\lambda)$ is thus obtained by taking the inverse value of 1.48 times a rolling MAD with the same window size. We see in Fig.~\ref{FigNoiseEx2}, that in the case of spectra that are limited by photon noise, our estimation of \snr{} using a rolling MAD on the YARARA river diagram maps (green curve) gives similar results as the theoretical value (red curve). However, in the case of spectra for which the noise is flat-field dominated, our estimation of \snr{} saturates at $\sim650$ (blue curve) while the theoretical \snr{} value due to photon-noise only is much higher (orange curve).

\begin{figure*}[t]
        \centering
        \includegraphics[width=18cm]{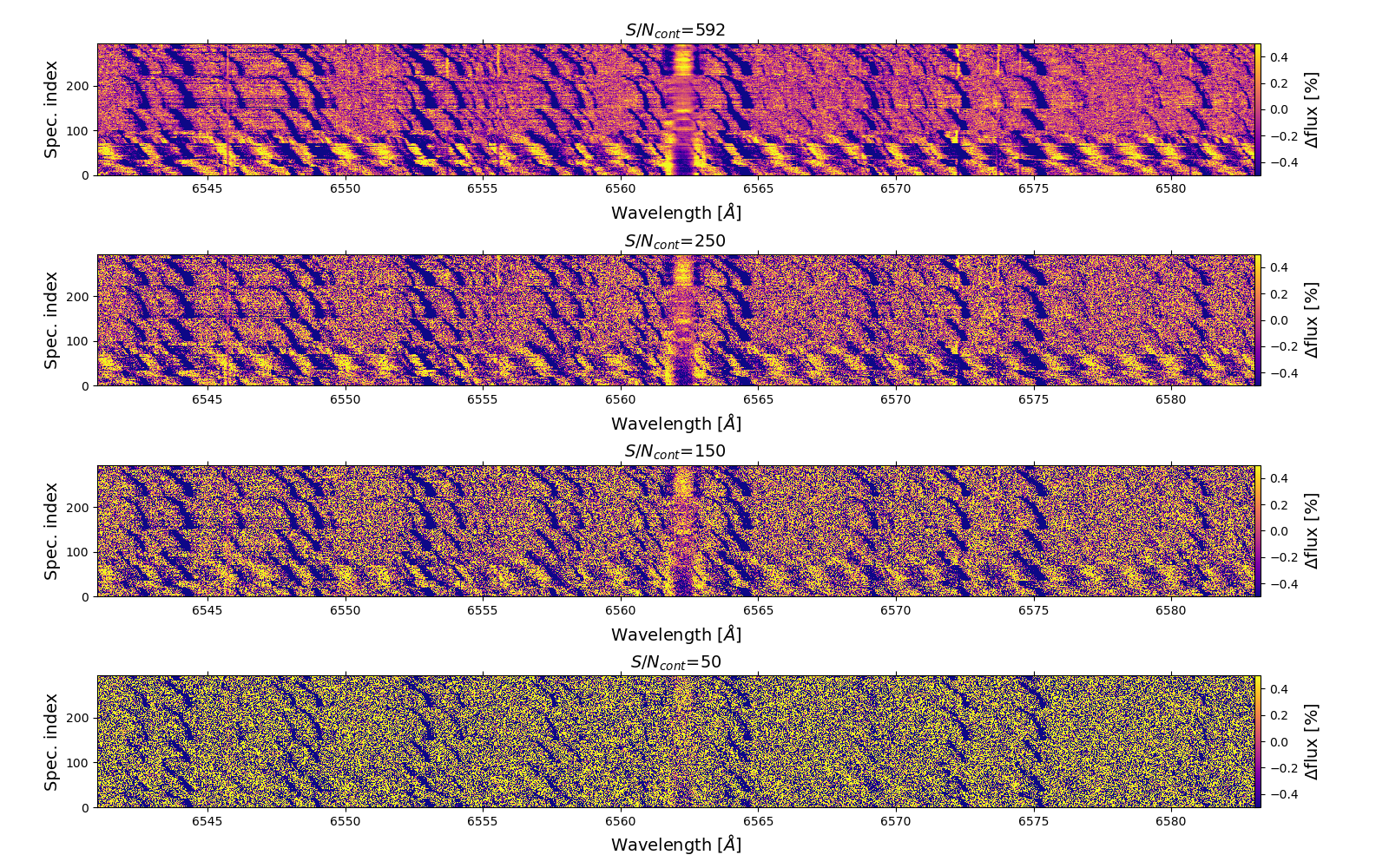}
        \caption{Examples of noise injections around the $H_\alpha=6562.79$ \ang{} line on the spectra time series of HD128621 (YARARA uncorrected) for different \snr{} values. From top to bottom: \snr{}$=592$ (nominal median values of the observations), 250, 150, and 50. This example also illustrates the approximated required \snr{} to detect and correct for the systematics using YARARA. For instance, at \snr{}=50, the pattern of interference and the H$_\alpha$ chromospheric variation are not detected whereas some of the telluric lines in that wavelength range are still visible.}
        \label{FigNoiseEx}
\end{figure*} 

In Sect.~\ref{sec:noise_sim}, we tested how our shell framework was able to mitigate stellar activity for different \snr{} level. To be able to obtain different \snr{} level from YARARA spectra, we note that lowering the \snr{} of a spectrum by a factor $x=($S/N$_{cont})/($S/N$_{cont}^{sim})$, implies an increase in the photon noise by a factor of $x$. As a YARARA spectrum already contains some noise $\sigma_{obs}$, to reach a photon noise $\sigma_{sim} = \sigma_{obs}\times x$ we only have to add as noise $\sigma_{add}$ to the spectrum:
\begin{eqnarray}
    \sigma_{add} &=& \sqrt{\sigma_{sim}^2 - \sigma_{obs}^2} = \sigma_{obs}\sqrt{x^2 - 1} \nonumber \\
    &=& \sigma_{obs}\sqrt{\left[\frac{S/N_{cont}}{S/N_{cont}^{sim}}\right]^2 - 1}.
\end{eqnarray}

We decided for the noise injections to produce 'white' photon noise, which means that the same S/N$_{cont}$ value was injected for all the wavelengths opposite to real observations on an instrument which are chromatic (see Fig.~\ref{FigNoiseEx2}). For all the noise injections, the same seed of the random generator was used and only the amplitude of the noise vector was changed. An example of the river diagram of HD128621 (systematics uncorrected by YARARA) is displayed in Fig.~\ref{FigNoiseEx} for three S/N$_{cont}^{sim}$ realisations (250, 150, and 50) around the H$_\alpha$ line.

\end{appendix}

\end{document}